\documentclass[]{jfm}
\usepackage{graphicx}
\usepackage{newtxtext}
\usepackage{newtxmath}
\usepackage{natbib}
\usepackage{color,xcolor}
\usepackage{hyperref}
\hypersetup{
    colorlinks = true,
    urlcolor   = blue,
    citecolor  = black,
}

\newcommand{\RomanNumeralCaps}[1]

\AtBeginDocument{
\captionsetup{
justification=justified,
width=\linewidth,
singlelinecheck=false
}
}
\makeatletter
\def\footerflagdefns#1{}
\def\ps@titlepage{
	\leftskip\z@\let\@mkboth\@gobbletwo\vfuzz=5\p@
	\def\@oddhead{
	}
	\def\@evenhead{\@j@urnal \hfil\llap{\thepage}}%
	\def\@oddfoot{\absfooterflag}%
	\def\@evenfoot{\absfooterflag}%
	\def\sectionmark##1{}%
	\def\subsectionmark##1{}%
}
\makeatother

\usepackage{overpic}
\newcommand{\s}{§}

\title{Formation of external particle jets on a spherical particle bed subjected to strong explosive loading}

\author{Yifeng He\aff{1,2}, Junsheng Zeng\aff{3,4}, Baolin Tian\aff{3,4}~\corresp{\email{tianbaolin@buaa.edu.cn}}, Yue Yang\aff{1,2}~\corresp{\email{yyg@pku.edu.cn}}}

\affiliation{
\aff{1}State Key Laboratory for Turbulence and Complex Systems, School of Mechanics and Engineering Science, Peking University, Beijing 100871, PR China
\aff{2}HEDPS-CAPT, Peking University, Beijing 100871, PR China
\aff{3}School of Aeronautic Science and Engineering, Beihang University, Beijing, PR China
\aff{4}Shanghai Zhangjiang Institute of Mathematics, Shanghai, PR China
}

\begin{document}
\maketitle

\begin{abstract}
We report the mechanism for the formation of external particle jets on a spherical particle bed subjected to strong explosive loading, revealing a critical dependence on particle size.
Under strong explosive loading, the formation of external particle jets is primarily driven by a drag-coupled mechanism.
We conducted Eulerian--Lagrangian simulations, with up to $2048^3$ effective cells and $1.8$ million tracked parcels on an adaptive mesh, for both small- and large-particle cases. 
Pronounced jets are observed only with small particles, alongside accelerated bed thickening. 
By defining characteristic inner and outer radii, the particle bed thickness evolution is quantified, showing an initial linear growth followed by a nonlinear deceleration.
Particle dynamics analysis indicates that drag force dominates particle motion and jet formation during the nonlinear stage.
The initial angular non-uniformity of the particle bed induces a non-uniform gas radial velocity. 
Through drag coupling, this flow asymmetry generates a radial velocity difference in small particles, thereby promoting pronounced jet formation, whereas large particles resist this drag-induced effect.
The greater drag-induced deceleration on smaller particles leads to an increased velocity difference across the particle bed, explaining the accelerated thickening.
A characteristic radius model that integrates the Gurney model for the linear stage with a drag-dominated deceleration model for the nonlinear stage is established and shows good agreement with numerical results across different particle sizes.

\end{abstract}


\section{Introduction}\label{sec:introduction}
High-speed multiphase flows occur in a wide range of natural phenomena and engineering processes, including volcanic eruptions~\citep{lube2020multiphase}, explosive combustion~\citep{zhang2001explosive,bala10}, detonations with reactive particles~\citep{Papalexandris2004}, dust dispersion~\citep{Lai2018} and plume-surface interactions~\citep{jesse2022}.
The explosive granular dispersal system represents a quintessential example of such flows. 
In this system, an explosive charge or the sudden release of high-pressure gas violently propelled outward densely packed particles to form particle clouds~\citep{xue23}.
Current research on explosive granular dispersal systems can be categorized into planar, cylindrical, and spherical geometries. 

The planar geometry is widely investigated using the interaction between shock waves and granular curtains within shock tubes. 
\citet{Rogue98} examined granular curtain thickness evolution following shock impingement using vertical shock tube experiments. 
These studies subsequently served as benchmarks for validating numerical simulations of high-speed gas-particle flows~\citep{McGrace2016,Tian20}. 
Later, \cite{wagner2012shock,theofanous2016dynamics,demauro2017unsteady} have employed gravity-fed particle curtains to probe the temporal evolution of particle curtain thickness. 
Based on experimental results, \citet{demauro2019improved} and \citet{daniel2022shock} derived scaling laws through force-balance analysis.
Furthermore, shock-induced interfacial instabilities of granular curtain in the planar configuration have been studied by numerical methods~\citep{chiapolino2020numerical,li2022shock,young2025effect}.

For cylindrical geometry, \citet{Rodriguez2013,rodriguez2014} investigated the explosively dispersed granular systems using a Hele-Shaw apparatus, clearly visualizing interfacial instabilities on the particle bed.
The formation of distinct internal and external particle jets on the internal and external surfaces of a particle bed, respectively, was observed in experiments~\citep{rodriguez2017,xue2018dual}.
Following initial shock impingement on the inner interface of the particle bed, internal particle jet develops. Later, external particle jetting forms from the outer interface of the particle bed.

The formation of external particle jets within cylindrical explosive-driven particle dispersal systems has attracted significant theoretical research interest~\citep{annamalai2016effects,osnes2018numerical,chiapolino2020numerical}. 
\citet{xue2018dual} observed that the external particle jet formation is a manifestation of Rayleigh-Taylor instability (RTI), whereas the internal jet cannot be classified as such due to its considerably delayed initiation and slow initial growth.
\citet{koneru2020numerical} showed that the deposition of vorticity through a gas-particle analogue of Richtmyer-Meshkov instability (RMI) plays a pivotal role in the channeling of particles into well-defined jets at the outer edge of the particle bed.
\citet{He2025} concluded that the RTI, the RMI, and large particle inertia contribute to the formation of the external particle jets, and proposed an external particle jet length model.

Both the aforementioned experimental and numerical studies fall within the regime of weak shock waves. 
Conversely, experimental investigations concerning explosive dispersal systems with spherical geometries typically employ high-energy explosives~\citep{frost2018heterogeneous}, generating strong shock waves.
\citet{zhang2001explosive} found that the detonation of a spherical charge produces a blast wave and supersonic gas-solid flow wherein particles can penetrate the shock front beyond a critical size limit.
\citet{Pontalier2018} experimentally investigated in spherical geometry and revealed that blast wave attenuation from a high explosive surrounded by inert material depends primarily on the mitigant-to-explosive mass ratio, with material properties playing a secondary role.

Numerical studies on the explosive dispersal granular systems in spherical geometry remain few, primarily due to the significant computational cost.
\citet{bala10} employed the Eulerian-Lagrangian approach to investigate the flow-field following the detonation of a spherical TNT charge within an ambient, dilute cloud of aluminum particles.
\citet{posey2024three} applied the Eulerian-Eulerian method to numerically investigate the ignition and combustion dynamics of a loosely packed spherical shell of aluminum particles surrounding a spherical explosive charge.
\citet{farrukh2025particle} studied  particle and fluid time scales in a spherical multiphase blast flow, and demonstrates that the equilibrium Eulerian method provides an accurate and computationally efficient alternative to Eulerian-Eulerian approaches for explosive particle dispersal when Stokes numbers less than one, with particle size being the dominant controlling factor.

The formation of external particle jets in spherical experimental configurations is governed by multiple factors, including particle size, material stiffness, saturation with liquid, and the ratio of the powder to explosive mass.
\citet{marr2018suppression} demonstrated that jet formation tendency partially depends on particle material properties: brittle ceramics and soft, ductile metals are more susceptible, while materials with moderate hardness, high compressive strength, and high toughness are less prone.
The experiments conducted by \citet{goroshin2016measurement} demonstrated the formation of external particle jets even during the explosive dispersal of small size steel particles.
\citet{Loiseau2018} found that water saturation of a porous SiC particle bed suppresses the formation of large coherent jets typical of dry beds, leading instead to the generation of numerous fine jets.
\citet{milne2010dynamic} found that the number of particle jets systematically decreases with an increasing ratio of powder to explosive mass.

Experimental evidence in spherical geometry reveals that the mechanism of particle jets formation under strong explosion conditions is fundamentally distinct from that induced by weak shock waves.
\citet{milne2010dynamic} discovered that the growth rate of the external particle jet in the spherical configuration is considerably higher than that of the classical RTI/RMI.
\citet{Milne2014} later proposed a conceptual model that the strong shock-compacted particle bed undergoes tensile fracture by the release wave, producing large fragments whose radial motion, through the shedding of unconsolidated particles, forms particle jets.

However, the formation mechanism of external particle jets under strong explosive loading remains incompletely understood, primarily due to the formidable challenges in experimentally observing the internal structures within spherical explosive dispersal systems, a difficulty arising from the optical opacity of the particulate media.
Consequently, three-dimensional (3D) spherical numerical simulations provide insights into the complex gas-particle flows in such configurations.

In this study, we employed an Eulerian-Lagrangian approach integrated with adaptive mesh refinement~\citep{zhang2019}, simulating a 3D spherical explosive dispersal system at a maximum effective resolution of $2048^3$ cells with approximately $1.8$ million particle parcels. 
Futhermore, we elucidate the formation of external particle jets and the evolution of the particle bed thickness, through analysis of high-fidelity simulation data.

The paper is organized as follows.
In \s\,\ref{sec:simulation}, we describe the numerical setup of particle beds and high-pressure gas pocket in the spherical configuration. 
Subsequently, we analyzed the angular clustering of particles through correlation analysis, defined the characteristic radii, and quantitatively examined the evolution of the particle bed using particle dynamics in \s\,\ref{sec:results}. 
Then we develop a characteristic radii model in \s\,\ref{sec:model}. 
Some conclusions are drawn in \s\,\ref{sec:conclusion}. 

\section{Simulation overview}\label{sec:simulation}
\subsection{Governing equations and numerical methods}
We conducted numerical simulations for explosive granular dispersal systems using the compressible multiphase particle-in-cell~(CMP-PIC) method~\citep{Tian20}.
This method integrates the Eulerian-Lagrangian approach to solve the motions of particle and gas.
The two phases are fully coupled via volume fraction and momentum exchange terms, while the gas field surrounding the particle surfaces is not resolved~\citep{cap2013}.

In the CMP-PIC method, a parcel for the particle phase consists of multiple particles with the same physical properties.
The contribution of parcel $i$ to the weighted particle volume fraction is $\alpha_{p,i}=w_{i,c}V_{p,i}/V_{c}$, where $V_{p,i}$ denotes the volume of parcel $i$, $V_{c}$ the computational cell volume, and $w_{i,c}$ the distributed weight that parcel $i$ contributes to the particle volume fraction in each cell.
The volume fraction of particle in a cell is $\alpha_{p}=\sum_{i}\alpha_{p,i}$, and the volume fraction of gas is $\alpha_{f}=1-\alpha_{p}$.
The cubic spline kernel function is used for calculating the particle volume fraction and source
terms on the Eulerian grid points, and the fluid variables on Lagrangian parcels.

The CMP-PIC method tracks parcels in the Lagrangian framework and takes parcel-parcel interactions into account.
The motion of parcels is governed by
\begin{equation}
	\frac{\mathrm{D}\boldsymbol{x}_{p,i}}{\mathrm{D}t}=\boldsymbol{v}_{p,i}
\end{equation}
and
\begin{equation}\label{eq:ptcdym}
	\frac{\mathrm{D}\boldsymbol{v}_{p,i}}{\mathrm{D}t}=\boldsymbol{a}_{p,i}=\frac{1}{m_{p,i}}\sum_{j}\boldsymbol{F}_{C,ij}-\boldsymbol{S}_{p,i}.
\end{equation}
Here, $\boldsymbol{a}_{p,i}$, $m_{p,i}$, $\boldsymbol{x}_{p,i}$, and $\boldsymbol{v}_{p,i}$ denote the acceleration, mass, location, and velocity of parcel $i$, respectively.
The collision force
\begin{equation}\label{eq:col}
	\boldsymbol{F}_{C,ij}=k_{n,p}\boldsymbol{\delta}_{n}+\frac{2\ln \epsilon_{p}}{\sqrt{\pi^{2}+\ln^2\epsilon_{p}}}\sqrt{m_{p,i}k_{n,p}}\boldsymbol{u}_{n,ij}
\end{equation}
between parcels $i$ and $j$ is calculated by a soft-sphere model~\citep{Liu14},
where the stiffness $k_{n,p}$ and the restitution coefficient $\epsilon_{p}$ characterize parcel properties, and the overlap distance $\boldsymbol{\delta}_{n}$ with the normal velocity difference $\boldsymbol{u}_{n,ij}$ describes contact dynamics between parcels.

The source term of parcel $i$ applied by the gas phase is
\begin{equation}\label{eq:source}
	\boldsymbol{S}_{p,i}=\frac{1}{\rho_{p}}\nabla p_{f}+D_{p,i}\left(\boldsymbol{v}_{p,i}-\boldsymbol{u}_{f}\right),
\end{equation}
with the particle density $\rho_{p}$, gas pressure $p_{f}$, and gas velocity $\boldsymbol{u}_{f}$. 
Here, the drag force coefficient
\begin{equation}\label{eq:drag}
D_{p,i}=C_d\frac{3\rho_f\left|\boldsymbol{v}_{p,i}-\boldsymbol{u}_f\right|}{4\rho_pd_p}
\end{equation}
is calculated using the Di Felice model combined with Ergun’s equation \citep{crowe1998}, 
with the dimensionless drag coefficient
\begin{equation}\label{eq:cd}
C_{d}=\frac{24}{\mathrm{Re}_{p}}\left\{\begin{array}{ll}
8.33 \frac{\alpha_{p}}{\alpha_{f}}+0.0972 \mathrm{Re}_{p}, &\alpha_{f}<0.8, \\
f_{\text {base }} \alpha_{f}^{-\zeta}, &\alpha_{f} \geq 0.8,
\end{array}\right.
\end{equation}
standard drag factor~\citep{McGrace2016}
\begin{equation}\label{eq:fbase}
f_{\text {base }}=\left\{\begin{array}{cc}
1+0.167 \mathrm{Re}_{p}^{0.687}, &\mathrm{Re}_{p}<1000, \\
0.0183 \mathrm{Re}_{p}, &\mathrm{Re}_{p} \geq 1000,
\end{array}\right.
\end{equation}
and 
the exponent
\begin{equation}
\zeta=3.7-0.65 \exp \left[-\frac{1}{2}\left(1.5-\log _{10} \mathrm{Re}_{p}\right)^{2}\right].
\end{equation}
Here, $d_{p}$ is the particle diameter, $\rho_{f}$ the gas density, $\mu_{f}$ the gas viscosity with the Sutherland model, and $\mathrm{Re}_{p}=\rho_{f}d_{p}\left|\boldsymbol{v}_{p,i}-\boldsymbol{u}_{f}\right|/\mu_{f}$ the particle Reynolds number.  
Note that specifying the valid range for the drag models is non-trivial~\citep{Osnes2023}.

The inertial properties of particles in a fluid flow is characterized by the Stokes number
\begin{equation}\label{eq:stokes}
 \mathrm{St}=\tau_{p}/\tau_{f}, 
\end{equation}
where the particle response time
\begin{equation}\label{eq:taup}
    \tau_{p}=\frac{\rho_{p}d_{p}^2}{18\mu_{f}\left(1+0.15\mathrm{Re}_{p}^{0.687}\right)}
\end{equation}
is evaluated using the Schiller-Naumann correction~\citep{berk2021}, and the flow field characteristic time
\begin{equation}\label{eq:tauf}
\tau_{f}=\frac{L_{\tau}}{V_{\tau}}
\end{equation}
is the ratio of the characteristic length $L_{\tau}$ to the characteristic velocity $V_{\tau}$ in the flow field. 

The evolution of parcel temperature is~\citep{crowe1998}
\begin{equation}
    m_{p,i}C_{p,p}\frac{\mathrm{d}T_{p,i}}{\mathrm{d}t}=-Q_{i},
\end{equation}
where $T_{p,i}$ is the temperature of the parcel $i$ and $C_{p,p}$ is the particle heat capacity and the interphase heat exchange term is given by
\begin{equation}\label{eq:heati}
    Q_{i}=\pi d_{p}k_{f}\mathrm{Nu}\left(T_{p,i}-T_{f}\right),
\end{equation}
where $k_{f}$ is the thermal conductivity of the gas phase, and $T_{f}$ denotes the gas temperature. 
The Nusselt number in  ~\eqref{eq:heati} is defined according to the correlation by~\citet{ranzmarshall1952}
\begin{equation}\label{eq:heatnu}
\mathrm{Nu}=2+0.6\mathrm{Re}_{p}^{1/2}\mathrm{Pr}^{1/3},
\end{equation}
where $\mathrm{Pr}=\mu_{f}C_{p,f}/k_{f}$ is the Prandtl number with the gas heat capacity $C_{p,f}$ at constant pressure.

The gas phase in CMP-PIC is discretized in the Eulerian framework using four-way coupling~\citep{wang2020four}. This approach extends the two-way coupling method~\citep{dalla2018clustering} by additionally accounting for particle-particle collisions and the influence of particle volume fraction on the gas, while maintaining the momentum and energy transfers between gas and particles.
The gas-phase governing equations~\citep{shall2020}
\begin{equation}
	\frac{\mathrm{D} \left(\alpha_{f}\rho_{f}\right)}{\mathrm{D}t}=-\alpha_{f}\rho_{f}\left(\nabla\cdot\boldsymbol{u}_{f}\right),
\end{equation}
\begin{equation}\label{eq:gasmom}
	\frac{\mathrm{D} \left(\alpha_{f}\rho_{f}\boldsymbol{u}_{f}\right)}{\mathrm{D}t}=-\alpha_{f}\rho_{f}\boldsymbol{u}_{f}\left(\nabla\cdot\boldsymbol{u}_{f}\right)-\alpha_{f}\nabla p_{f}+\sum_{i}\rho_{p,i}\alpha_{p,i}\boldsymbol{S}_{p,i},
\end{equation}
\begin{equation}\label{eq:energy}
\frac{\mathrm{D}\left(\alpha_{f}\rho_{f}E_{f}\right)}{\mathrm{D}t}=-\alpha_{f}\rho_{f}E_{f}\left(\nabla\cdot\boldsymbol{u}_{f}\right)-\nabla\cdot\left(\alpha_{f}p_{f}\boldsymbol{u}_{f}\right)+\sum_{i}\rho_{p,i}\alpha_{p,i}\boldsymbol{S}_{p,i}\cdot\boldsymbol{v}_{p,i}+\mathcal{Q},
\end{equation}
are simplified from the model of \cite{BN86}, where $\rho_{p,i}$ is the density of parcel $i$, the subscript $i$ denotes the parcel $i$, $E_{f}=e+\left(\boldsymbol{u_{f}}\cdot\boldsymbol{u_{f}}\right)/2$ denotes the total energy of gas per unit volume, and $\mathcal{Q}=\sum_{i}\rho_{p,i}\alpha_{p,i}Q_{i}$ the heat exchange term.
They are closed with the equation of state for ideal gas
\begin{equation}\label{eq:eos}
	p_{f}=\rho_{f}\frac{\mathcal{R}}{M}T_{f}=\rho_{f} e\left(\gamma-1\right),
\end{equation}
with the gas density $\rho_{f}$, ideal gas constant $\mathcal{R}$, molar mass $M$, specific heat ratio $\gamma=C_{p,f}/C_{v,f}$, and internal energy $e$.
Here $C_{v,f}$ is the gas specific heat at constant volume.

The CMP-PIC has been validated against several benchmark experiments involving shock driven particle laden flows~\citep{Tian20}, including Rogue’s experiments~\citep{Rogue98}, the experiments conducted by \citet{Britan06}, the cavity formation in granular media due to explosions~\citep{Zeng2024} and the experiments of shock induced interfacial instability of granular media~\citep{xue2018dual}.

\subsection{Simulation setup}\label{subsec:setup}
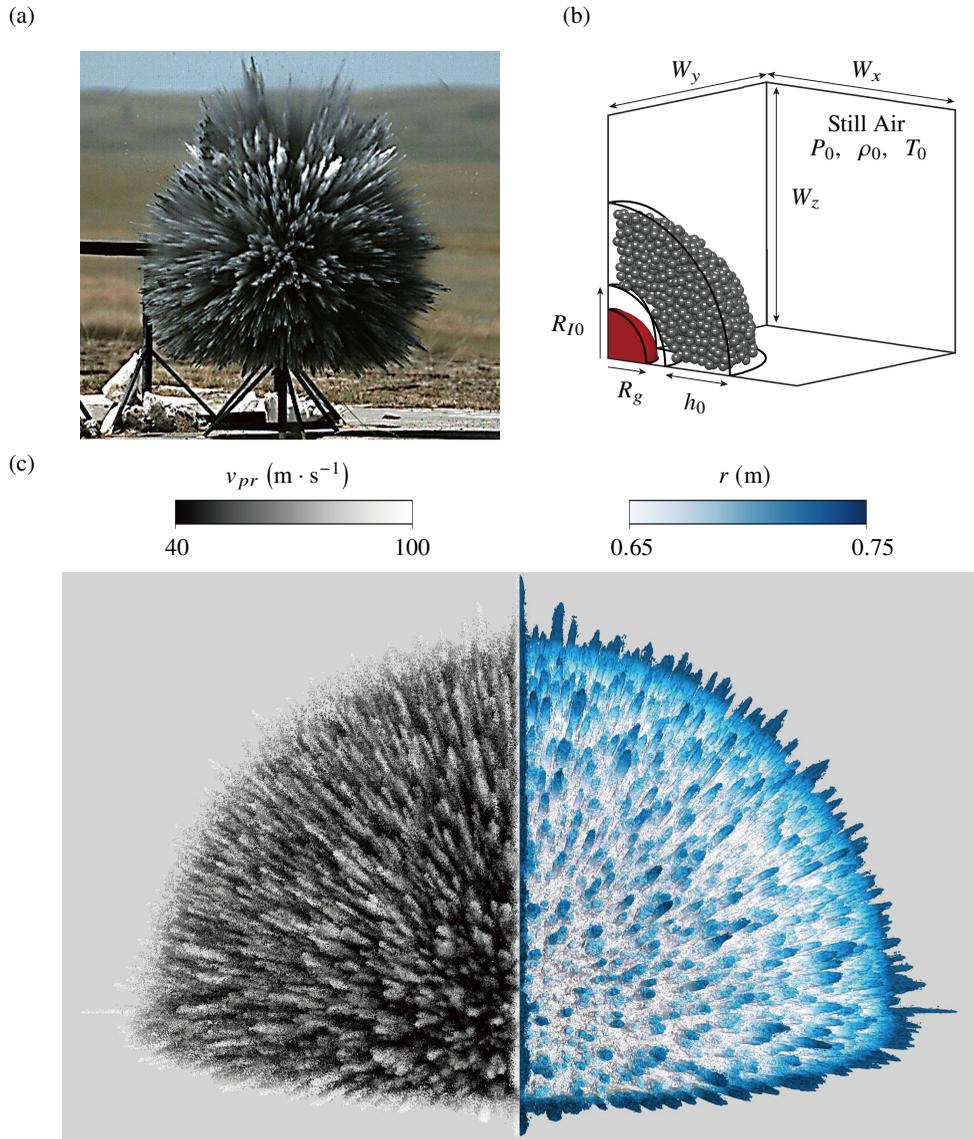
\begin{figure}
	\centering
     \begin{overpic}
	     [width=1.0\textwidth]{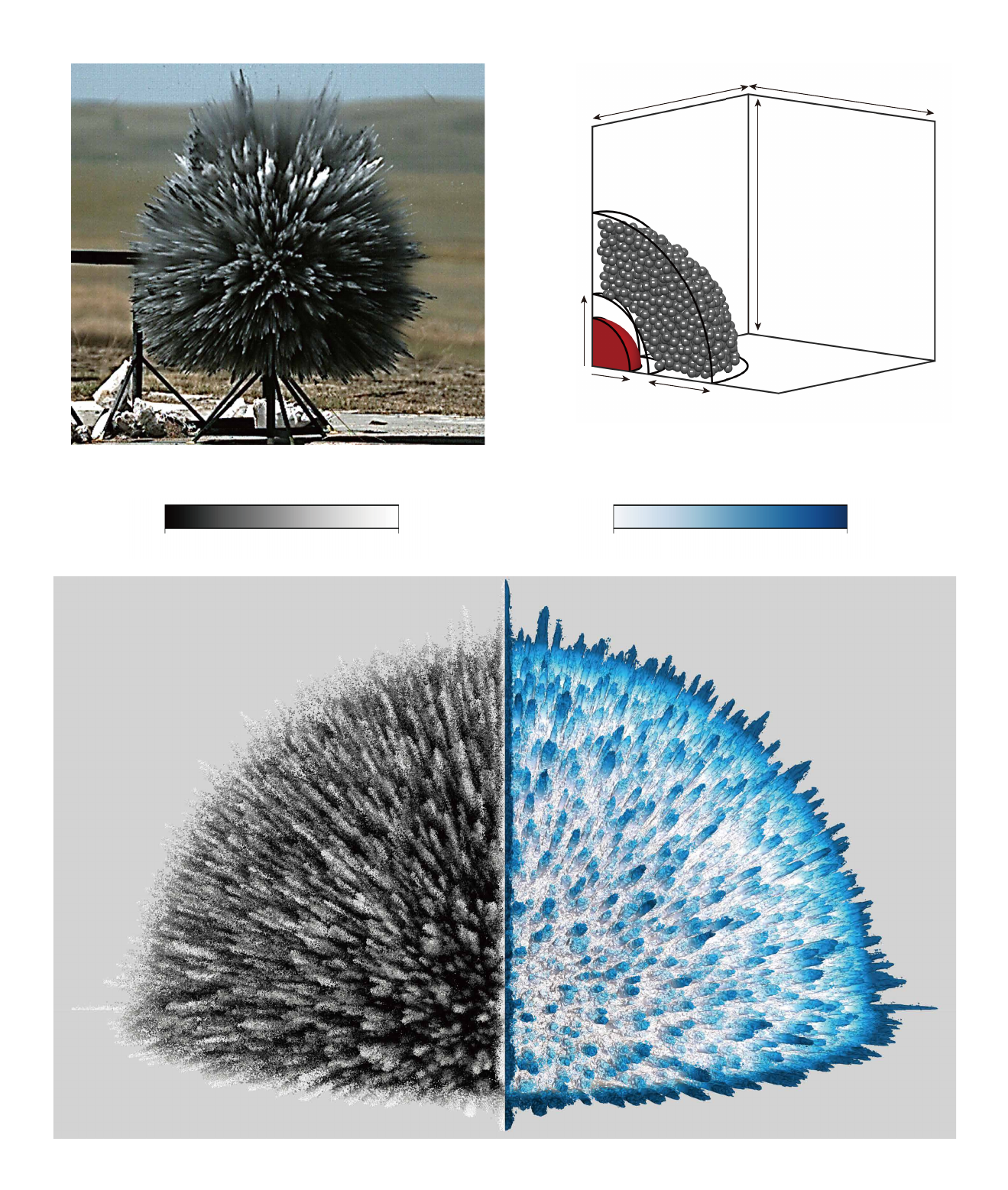}
        \put(0,97){\footnotesize{(a)}}
        \put(45,71.5){\footnotesize{$R_{I0}$}}
        \put(50.5,66){\footnotesize{$R_{g}$}}
        \put(56,65){\footnotesize{$h_{0}$}}
        \put(55,92.5){\footnotesize{$W_{y}$}}
        \put(65,82){\footnotesize{$W_{z}$}}
        \put(70,92.5){\footnotesize{$W_{x}$}}
        \put(46,97) {\footnotesize{(b)}}
        \put(0,60) {\footnotesize{(c)}}
        \put(68,88){\footnotesize{Still Air}}
	    \put(66.5,86){\footnotesize{$P_{0},$}}
		\put(70.5,86){\footnotesize{$\rho_{0},$}}
		\put(74.5,86){\footnotesize{$T_{0}$}}
        \put(18,59){\footnotesize{$v_{pr}\left(\mathrm{m\cdot s^{-1}}\right)$}}
        \put(13,53){\footnotesize{$40$}}
        \put(32,53){\footnotesize{$100$}}    \put(59,59){\footnotesize{$r\left(\mathrm{m}\right)$}}
        \put(50,53){\footnotesize{$0.65$}}
        \put(70,53){\footnotesize{$0.75$}}  
	\end{overpic}
\caption{
(a)~Dispersal of FE-110 particles with average sizes $d_{p}=32.1~\mathrm{\upmu m}$~\citep{goroshin2016measurement}. 
Image courtesy of David Frost, private communication, 2025.
(b)~Schematic of the 3D explosive dispersal system set-up in the numerical simulation.
(c)~Numerical results for case DP50 at time $t = 5\mathrm{ms}$. 
The left panel displays the parcel distribution, colored by the parcel radial velocity, while the right panel shows the iso-surfaces of $|\boldsymbol\omega_{f}|=3\sqrt{\langle\Omega\rangle}$, mapped with the radial coordinate.
Here $\Omega =|\boldsymbol\omega_{f}|^{2}/2$ denotes the enstrophy and $\langle\cdot\rangle$ represents the volume average over $\mathcal{D}$.
Note that the visualized domain consists of two mirrored $1/8$-spherical sections, whereas the actual computation was performed only on a single $1/8$-spherical domain.
}\label{fig:setup_scheme}	
\end{figure}

To investigate the particle jetting phenomenon observed in explosive dispersal of granular materials in figure~\ref{fig:setup_scheme}(a), we conducted a 3D numerical simulation of a spherical system. As shown in figure~\ref{fig:setup_scheme}(b), a high-pressure gas pocket with radius $R_{g}$ expands the surrounding particle bed with initial inner radius $R_{I0}$ and thickness $h_{0}$. 
Thus, the initial outer radius of the particle bed is $R_{O0}=R_{I0}+h_{0}$.
This problem is simulated in the domain $\mathcal{D}=\left[0,W_{x}\right]\times\left[0,W_{y}\right]\times\left[0,W_{z}\right]$, with domain sizes $W_x$, $W_y$ and $W_z$ in $x$-, $y$- and $z$-directions, respectively.
The computational domain is reduced to a $1/8$ sphere to leverage symmetry and minimize computational cost while retaining the essential physics.
Both the gas pocket and the particle bed are set at the origin $(x_c,y_c,z_c)=(0,0,0)$. 
The spherical coordinate $(r,\theta,\phi)$ is related to the Cartesian coordinate $(x,y,z)$ as $r=((x-x_c)^2+(y-y_c)^2+(z-z_c)^2)^{1/2}$, $\theta=\arctan(y-y_c,x-x_c)$ and $\phi=\arccos(z-y_c,r)$.

The high-pressure gas pocket is filled with still air with pressure $P_{s}$, density $\rho_{s}$ and temperature $T_{s}$.
For the still air, we set the initial ambient pressure $P_0=1.013\times10^5~\rm{Pa}$, temperature $T_0=299.5~\rm{K}$, and density $\rho_{0}=P_0M_{air}/\left(\mathcal{R}T_0\right)$ with the species molar mass $M_{air}=29~\mathrm{g}~\mathrm{mol^{-1}}$.
The radius, pressure and temperature of the gas pocket are set as $R_{g}=16~\mathrm{mm}$, $P_{s}=1.013\times10^9~\rm{Pa}$ and $T_{s}=299.5~\mathrm{K}$, respectively.
Thus, the equation of state for ideal gas in~\eqref{eq:eos} gives $\rho_s/\rho_0=P_s/P_0$.

The spherical-shell particle bed with initial volume fraction $\Phi_0=0.5$ is filled by random distributed computational parcels~\citep{Lozano16}.
The diameter of the parcel is set to avoid potential crystallization during shock compaction~\citep{xue23}.
The restitution coefficient $\epsilon_{p}=0.6$ and the normal stiffness  $k_{n,p}=2.0\times10^{7}~\mathrm{N\cdot m^{-1}}$ of contacts between parcels for calculating the particle collision force are set in~\eqref{eq:col}.
The constant initial inner radius $R_{I0}=20~\mathrm{mm}$ of the particle bed is set with the initial thickness $h_{0}=40~\mathrm{mm}$. 
The mass ratio between the surrounding particles and the gas pocket~\citep{xue23} is
\begin{equation}\label{eq:mass_ratio}
    \Gamma_M=\frac{\left(R_{O0}^3-R_{I0}^3\right)\Phi_0\rho_p}{ R_{g}^3\rho_{s}}.
\end{equation}
The thermal parameters $k_f = 0.026~\mathrm{W~m^{-1}~K^{-1}}$ and $C_{p,f} = 1.0~\mathrm{kJ~kg^{-1}~K^{-1}}$ in~\eqref{eq:heati} and~\eqref{eq:heatnu} are set, the Sutherland law
\begin{equation}
\mu_{f}=\mu_{f,r}\left(\frac{T_{f}}{T_{0}}\right)^{3/2}\left(\frac{T_{0}+T_{f,r}}{T_{f}+T_{f,r}}\right)
\end{equation}
is adopted to calculate the gas viscosity, with reference viscosity~$\mu_{f,r}=1.92\times10^{-5}~\mathrm{Pa\cdot s}$ and reference temperature~$T_{f,r}=110.56~\mathrm{K}$. 

This study comparatively examines two particle size regimes, referred to as case DP50 for small particles~($d_{p}=50~\mathrm{\upmu m}$) and case DP450 for large particles~($d_{p}=450~\mathrm{\upmu m}$), by using adaptive mesh refinement~\citep{zhang2019}. 
In case DP50, approximately $1.8$ million parcels were computed. 
The computational mesh employed a base grid with a resolution of $512^3$, which was refined by a factor of two across two levels, resulting in a finest grid resolution of $2048^3$ and a grid size of $\Delta x=0.5~\mathrm{mm}$.
For case DP450, approximately $0.54$ million parcels were tracked.
The coarse grid has a resolution of $256^3$. With two levels of refinement at a factor of two, the finest grid reaches a resolution of $1024^3$, corresponding to a grid size of $\Delta x=1.0~\mathrm{mm}$.
The parcel sizes are uniformly distributed from $285~\mathrm{\upmu m}$ to $474~\mathrm{\upmu m}$ in case DP50, and from $476~\mathrm{\upmu m}$ to $667~\mathrm{\upmu m}$ in case DP450. 
The time stepping $\Delta t$ is chosen to satisfy both the Courant--Friedrichs--Lewy condition and the time step limit of the discrete element method~\citep{Tian20}.
The third-order Runge--Kutta method is applied for the time integration.

The sound speed of air $c_{s}=\sqrt{\gamma P_{0}/\rho_{0}}$ and $h_0$ are used as the characteristic velocity $V_{\tau}$ and length $L_{\tau}$ of the flow field, respectively, to calculate the Stokes number.
Substituting $\mathrm{Re}_{p}=\rho_{0}d_{p}c_{s}/\mu_{f}$, \eqref{eq:tauf} and \eqref{eq:taup} into \eqref{eq:stokes} yields $\mathrm{Re}_{p}=9669$ and $\mathrm{St}=O\left(100\right)$ for case DP450, while for case DP50, the corresponding values are $\mathrm{Re}_{p}=1074$ and $\mathrm{St}=O\left(10\right)$, demonstrating that the Stokes number for the larger particles is an order of magnitude higher than that of the smaller particles.
The maximum particle Mach number $\mathrm{Ma}_{p}=\left|\boldsymbol{v}_{p,i}-\boldsymbol{u}_{f}\right|/c_{s}$ is around three in the present study.

The numerical results presented in figure~\ref{fig:setup_scheme}(c) demonstrate qualitative agreement with the experimental results shown in figure~\ref{fig:setup_scheme}(a), where external particle jet structures are prominently observed. 
The left panel of figure~\ref{fig:setup_scheme}(c) demonstrates that the radial velocity of particles at the head of the external particle jet exceeds that in the trailing region. 
Meanwhile, the right panel displays the isosurfaces of the magnitude $|\boldsymbol{\omega}_{f}|$ of vorticity $\boldsymbol{\omega}_{f}=\nabla\times\boldsymbol{u}_{f}$.
The numerical results also agree quantitatively with the experimental results, as elaborated in Appendix~\ref{app:cpwithex}.

\section{Simulation results}\label{sec:results}
\subsection{Evolution of the particle bed}
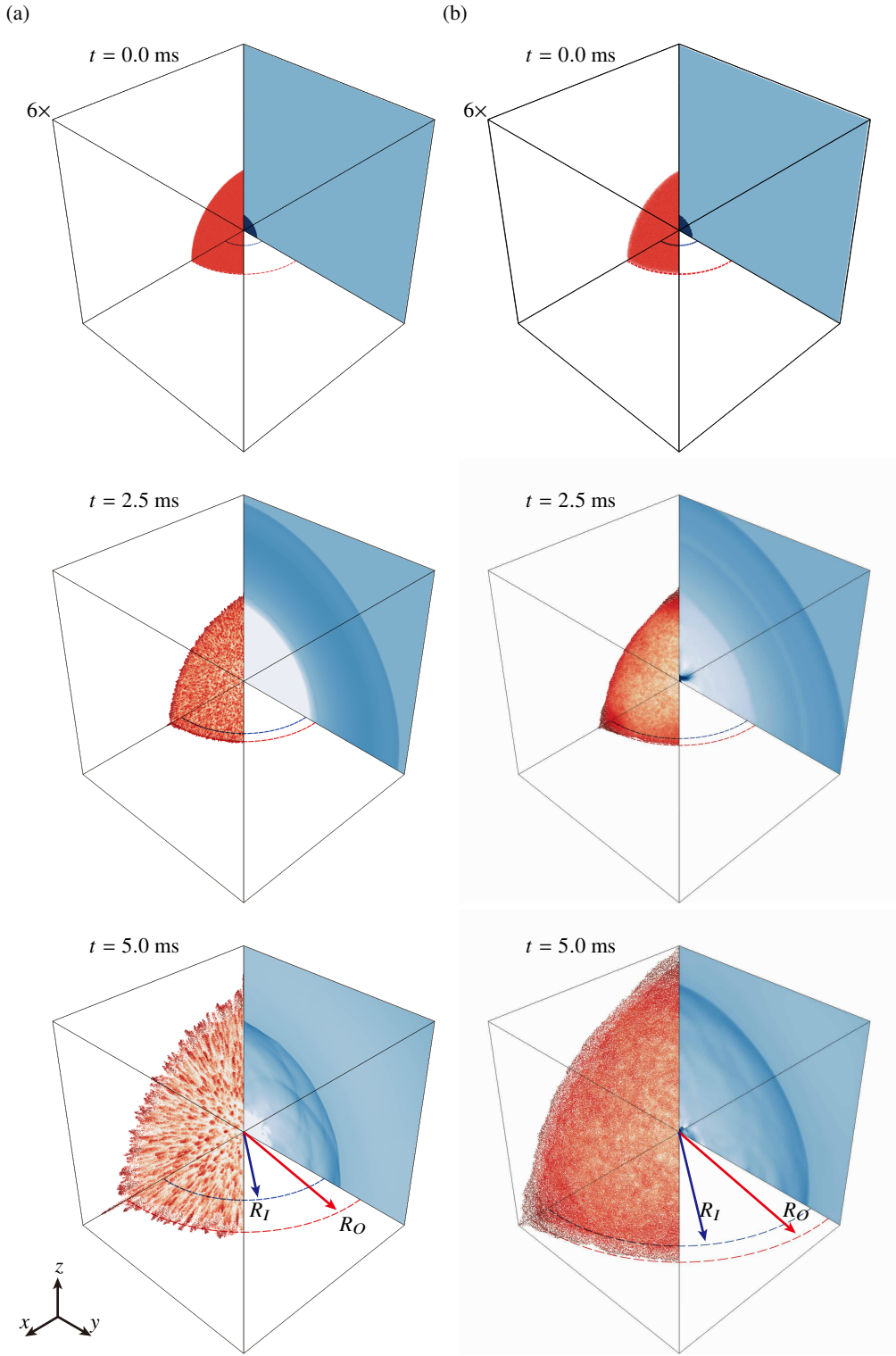
\begin{figure}
	\centering
     \begin{overpic}
	     [width=1.0\textwidth]{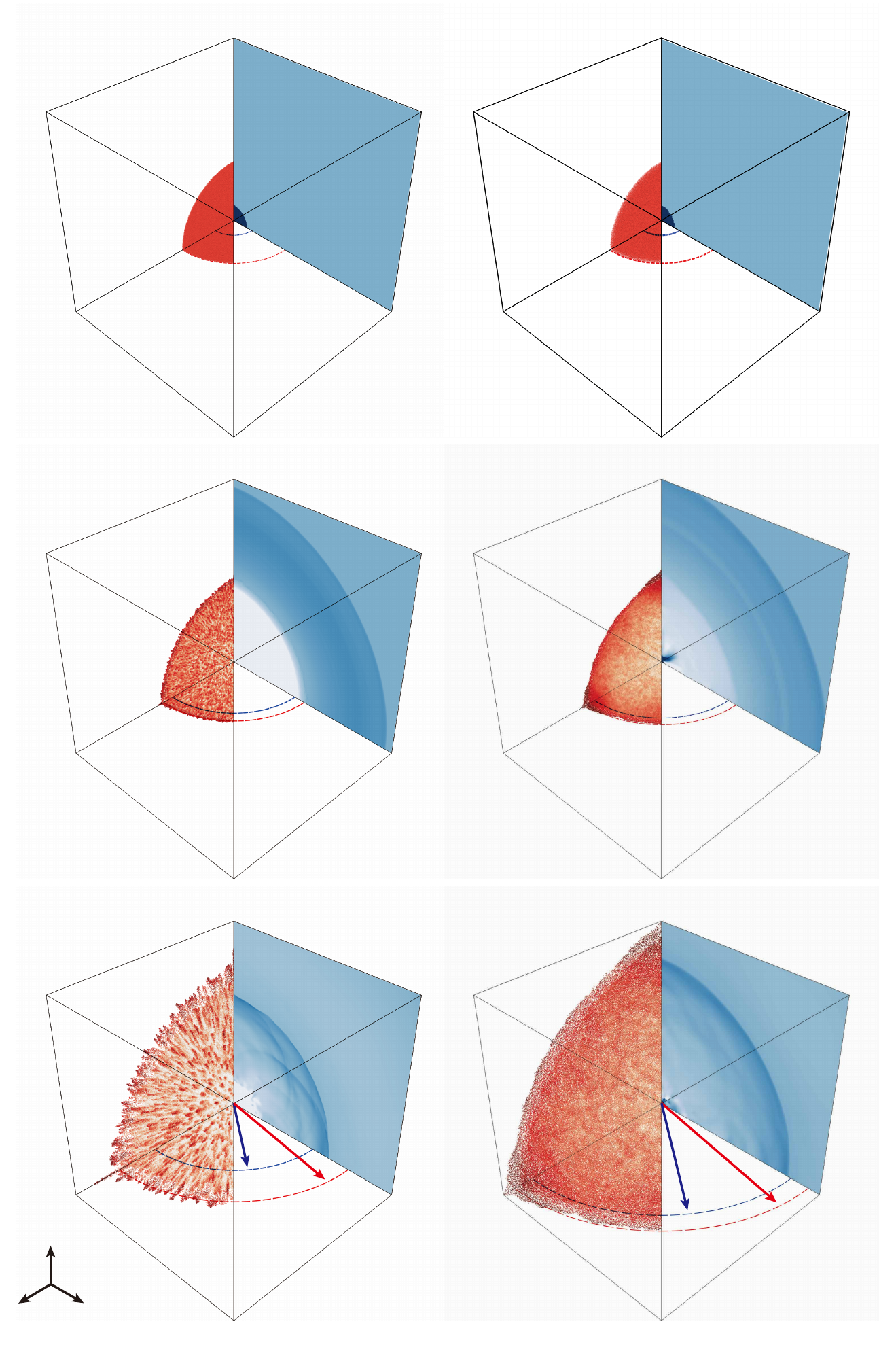}
        \put(0,99){\footnotesize{(a)}}
        \put(31.5,99){\footnotesize{(b)}}
        \put(6,96){\footnotesize{$t=0.0~\mathrm{ms}$}}
        \put(37.5,96){\footnotesize{$t=0.0~\mathrm{ms}$}}
        \put(6,64){\footnotesize{$t=2.5~\mathrm{ms}$}}
        \put(37.5,64){\footnotesize{$t=2.5~\mathrm{ms}$}}
        \put(6,32){\footnotesize{$t=5.0~\mathrm{ms}$}}
        \put(37.5,32){\footnotesize{$t=5.0~\mathrm{ms}$}}
        \put(1.5,92){\footnotesize{$6\times$}} 
         \put(33,92)
        {\footnotesize{$6\times$}}  
        \put(1,5)
        {\footnotesize{$x$}}  
        \put(6,5)
        {\footnotesize{$y$}}  
        \put(3.5,9)
        {\footnotesize{$z$}}  
         \put(17.5,13)
        {\footnotesize{$R_{I}$}} 
        \put(24,12)
        {\footnotesize{$R_{O}$}} 
        \put(50,13)
        {\footnotesize{$R_{I}$}} 
        \put(56,13)
        {\footnotesize{$R_{O}$}} 
	\end{overpic}
\caption{%
Evolution of parcel distributions, characteristic radii and pressure fields in cases (a)~DP50 and (b)~DP450.
Parcels are colored by the parcel radial velocity, with the inner radius $R_{I}$ and outer radius $R_{O}$ marked by blue and red dashed lines, respectively. The pressure contour is color-coded from white to blue within the range from $0.7$ to $1.4~\mathrm{bar}$.
For clarity, the images at $t = 0~\mathrm{ms}$ are magnified six times.
}\label{fig:flow_field}	
\end{figure}

The contour plot of gas pressure and position of the computational parcels are shown in figure~\ref{fig:flow_field}. 
Following the impact and acceleration by high-pressure gas in the central region after $t = 0~\mathrm{ms}$, the shell particle bed in both cases DP50 and DP450 undergoes expansion. 
By $t = 2.5~\mathrm{ms}$, the pressure in the central zone of each case drops below atmospheric pressure, leading to particle deceleration.  
Notably, small-scale particle jets emerge in case DP50, whereas no such structure is observed in case DP450.  
The particles in case DP50 experience more strong deceleration compared to those in case DP450. 
Subsequently, the initially small-scale jets in case DP50 evolves into distinct particle jets at $t = 5.0~\mathrm{ms}$, while case DP450 still exhibits no obvious jet formation

To quantify the evolution of the particle bed, the inner and outer radii $R_{I}$ and $R_{O}$ are defined as the mean radial positions of the innermost and outermost one percent of parcels, respectively. 
The dashed lines in figure~\ref{fig:flow_field} demonstrate that $R_{I}$ and $R_{O}$ capture the structural evolution of the particle bed.

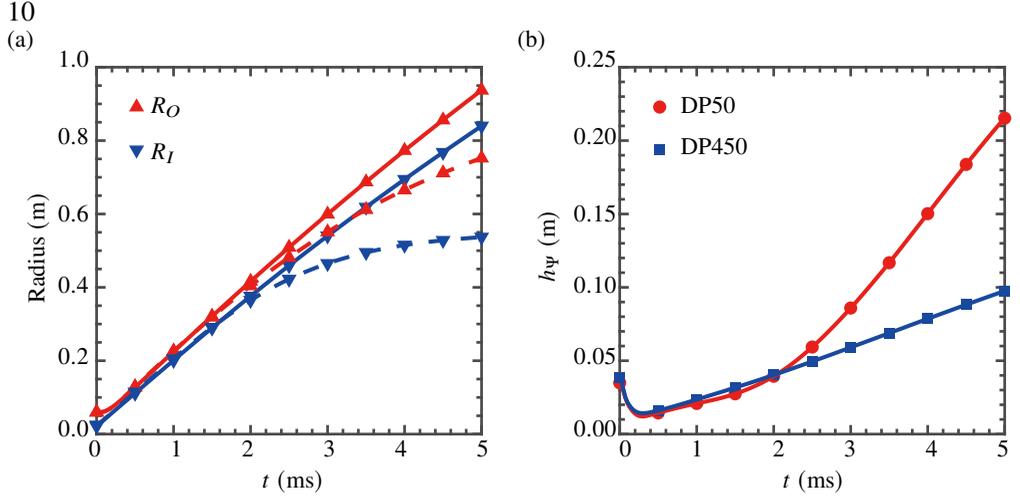
\begin{figure}
	\centering
     \begin{overpic}
	     [width=1.0\textwidth]{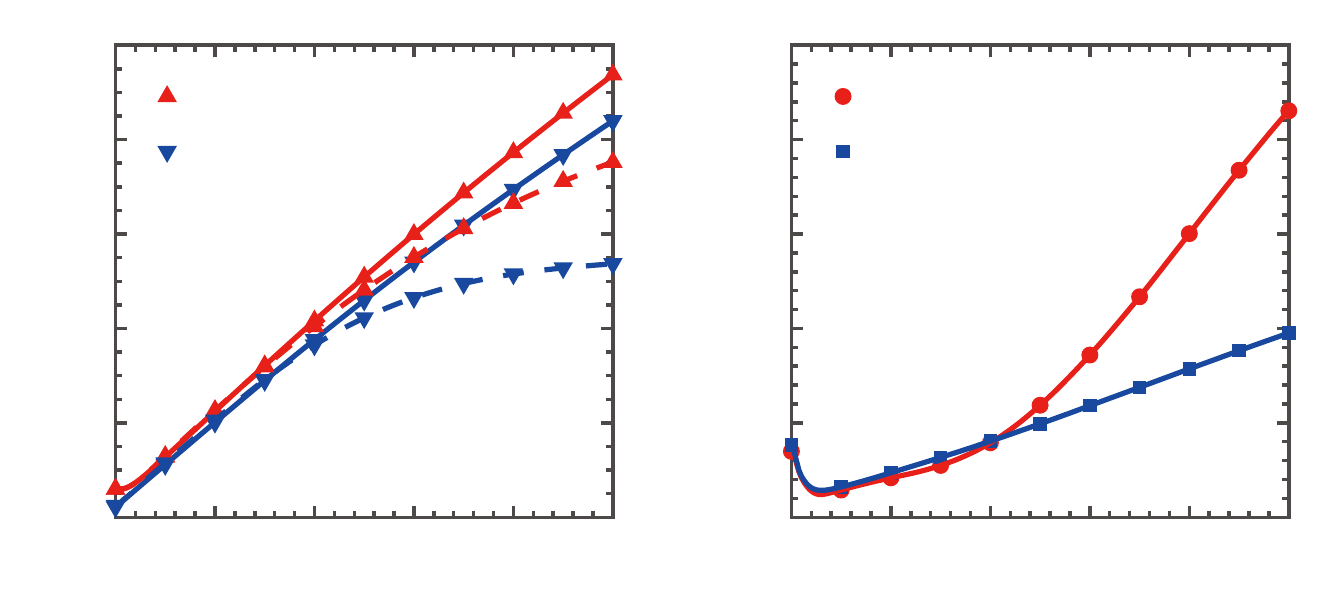}
      \put(14,38.5){\footnotesize{$R_{O}$}}
      \put(14,34){\footnotesize{$R_{I}$}}
      \put(66,38.5){\footnotesize{DP50}}
      \put(66,34.5){\footnotesize{DP450}}
      \put(0,45){\footnotesize{(a)}}
      \put(50,45){\footnotesize{(b)}}
      \put(8,5){\footnotesize{$0$}}
      \put(16,5){\footnotesize{$1$}}
      \put(23,5){\footnotesize{$2$}}
      \put(31,5){\footnotesize{$3$}}
      \put(38,5){\footnotesize{$4$}}
      \put(46,5){\footnotesize{$5$}}
      \put(60,5){\footnotesize{$0$}}
      \put(67,5){\footnotesize{$1$}}
      \put(75,5){\footnotesize{$2$}}
      \put(82,5){\footnotesize{$3$}}
      \put(90,5){\footnotesize{$4$}}
      \put(97,5){\footnotesize{$5$}}
     \put(5,7){\footnotesize{$0.0$}}
     \put(5,14){\footnotesize{$0.2$}}
     \put(5,21){\footnotesize{$0.4$}}
     \put(5,28){\footnotesize{$0.6$}}
     \put(5,35.5){\footnotesize{$0.8$}}
     \put(5,43){\footnotesize{$1.0$}}
     \put(55.5,7){\footnotesize{$0.00$}}
     \put(55.5,14){\footnotesize{$0.05$}}
     \put(55.5,21){\footnotesize{$0.10$}}
     \put(55.5,28){\footnotesize{$0.15$}}
     \put(55.5,35.5){\footnotesize{$0.20$}}
     \put(55.5,43){\footnotesize{$0.25$}}
      \put(25,2){\footnotesize{$t~(\mathrm{ms})$}}
     \put(76,2){\footnotesize{$t~(\mathrm{ms})$}}
      \put(2,20){\rotatebox{90}{\footnotesize{$\text{Radius}~(\mathrm{m})$}}}
     \put(52,22){\rotatebox{90}{\footnotesize{$h_{\Psi}~(\mathrm{m})$}}}
      \end{overpic}
\caption{%
(a)~Evolution of characteristic radii in cases DP50 (dashed lines) and DP450 (solid lines).
(b)~Evolution of the particle bed thickness.
}\label{fig:charac_radii}	
\end{figure}

Figure~\ref{fig:charac_radii}(a) plots the evolution of the characteristic radii. 
When high-pressure gases are released from the gas pocket, an incident shock wave impinges on the inner surface of the particle bed, resulting in a linear growth of $R_{I}$.
After the compaction front reaches the outer surface of the particle bed around $t=0.2~\mathrm{ms}$, $R_{O}$ starts to grow.
For both cases DP50 and DP450, the growth of $R_{I}$ undergoes stronger deceleration than that of $R_{O}$.
This difference arises because the gas-particle velocity difference is larger near $R_{I}$ than in the vicinity of $R_{O}$, leading to stronger local gas-particle coupling forces acting on the particles close to $R_{I}$~(see details in \S\,\ref{subsec:effects}). 
In case DP50, $R_{O}$ and $R_{I}$ experience a short period of linear growth, and the growth is mitigated around $t=3~\mathrm{ms}$ due to strong inward gas-particle coupling forces.
In contrast, for case DP450, $R_{O}$ and $R_{I}$ keep nearly linear growth due to large particle inertial.

The particle bed thickness
\begin{equation}
    h_{\Psi}=R_{O}-R_{I},~~~\Psi=l,s
\end{equation}
is defined as the radius difference between the inner side and outer side of the particle bed, where subscripts $l$ and $s$ denote cases DP50 and DP450, respectively.
The evolution of $h_{l}$ and $h_{s}$ is shown in figure~\ref{fig:charac_radii}(b).
Both the cases with large and small $d_{p}$ undergo a compression stage until around $t=0.2~\mathrm{ms}$, when the particle bed thickness reaches its minimum value.
Before $t=2~\mathrm{ms}$, both $h_{s}$ and $h_{l}$ exhibit a nearly identical and approximately linear increase.
When $t>2~\mathrm{ms}$, however, $h_{s}$ develops non-linearly, accelerating significantly, whereas $h_{l}$ continues to follow its roughly linear growth.
This growth disparity originates from the significantly lower inertia of small particles compared to large particles, resulting in enhanced gas-particle coupling.
The detailed physical mechanisms will be elucidated in \S\,\ref{subsec:effects}.

\subsection{Formation of external particle jets}\label{subsec:effects}
The angular clustering analysis, conducted using the box-counting method in Appendix~\ref{app:angular}, reveals only a weak correlation between particle clustering and jet formation.
This finding stands in contrast to mechanisms driven by weak shock waves, where vorticity deposition generates strong angular clustering that defines the jet pattern~\citep{koneru2020numerical}.
Hence, the distinct nature of the jet formation mechanism under strong explosive loading necessitates further analysis.

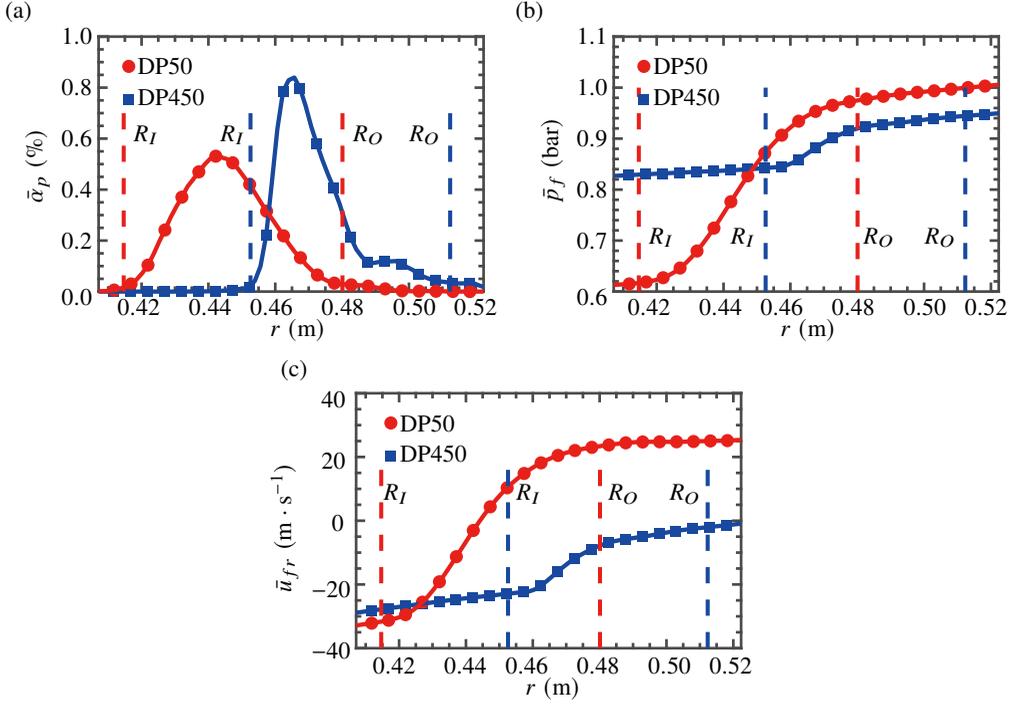
\begin{figure}
	\centering
     \begin{overpic}
	     [width=1.0\textwidth]{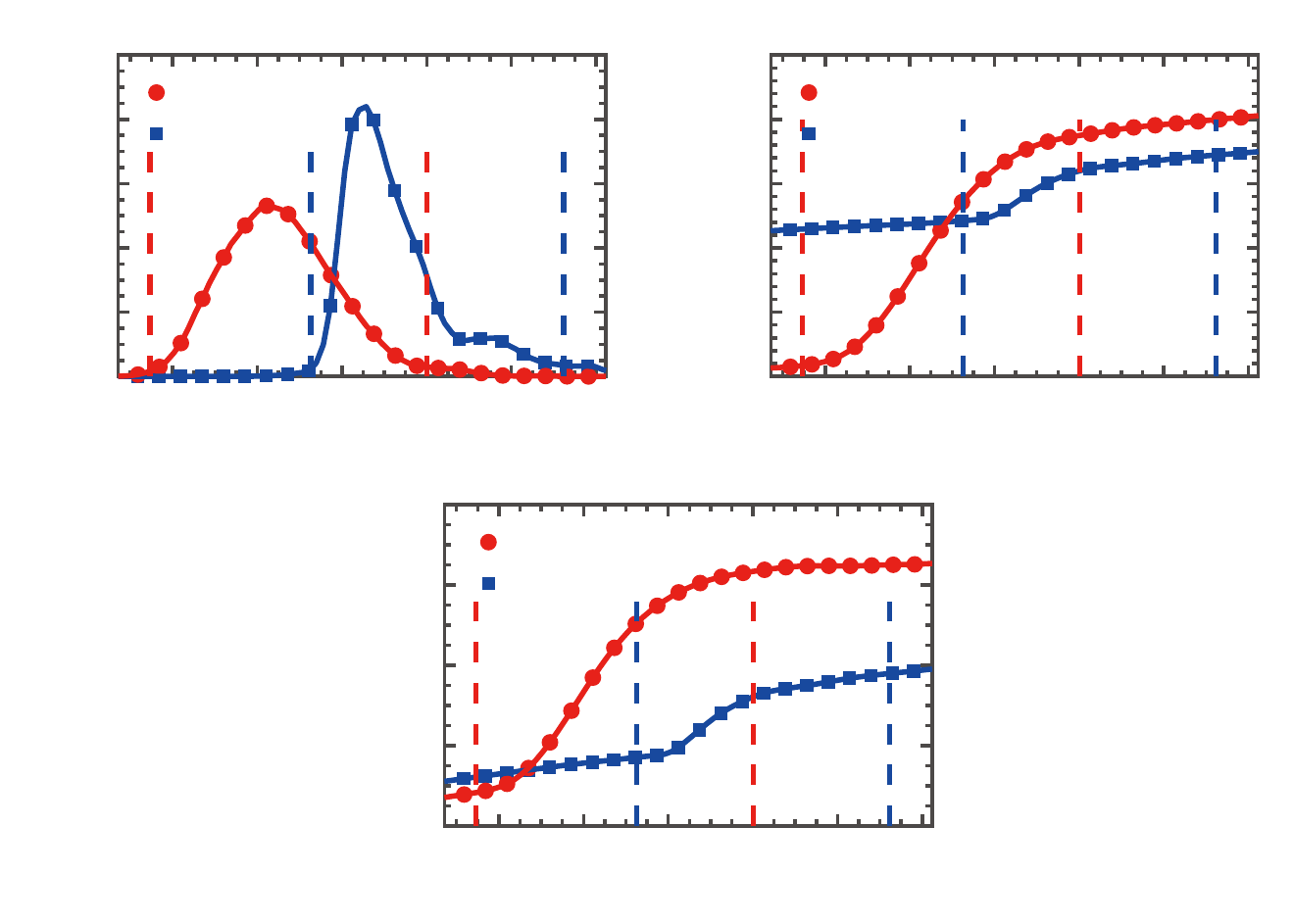}
      \put(2,51){\rotatebox{90}{\footnotesize{$\bar{\alpha}_{p}~(\%)$}}}
\put(5.5,41){\footnotesize{$0.0$}}
\put(5.5,46.5){\footnotesize{$0.2$}}
\put(5.5,51.5){\footnotesize{$0.4$}}
\put(5.5,56.5){\footnotesize{$0.6$}}
\put(5.5,61.5){\footnotesize{$0.8$}}
\put(5.5,66.5){\footnotesize{$1.0$}}
      \put(52.5,51){\rotatebox{90}{\footnotesize{$\bar{p}_{f}~(\mathrm{bar})$}}}
\put(56,41){\footnotesize{$0.6$}}
\put(56,46.5){\footnotesize{$0.7$}}
\put(56,51.5){\footnotesize{$0.8$}}
\put(56,56.5){\footnotesize{$0.9$}}
\put(56,61.5){\footnotesize{$1.0$}}
\put(56,66.5){\footnotesize{$1.1$}}
      \put(26,13){\rotatebox{90}{\footnotesize{$\bar{u}_{fr}~(\mathrm{m\cdot s^{-1}})$}}}
\put(30,6.5){\footnotesize{$-40$}}
\put(30,12.5){\footnotesize{$-20$}}
\put(32,19){\footnotesize{$0$}}
\put(31,25){\footnotesize{$20$}}
\put(31,31){\footnotesize{$40$}}
      \put(26,38){\footnotesize{$r~(\mathrm{m})$}}
\put(11,40){\footnotesize{$0.42$}}
\put(18,40){\footnotesize{$0.44$}}
\put(24,40){\footnotesize{$0.46$}}
\put(31,40){\footnotesize{$0.48$}}
\put(38,40){\footnotesize{$0.50$}}
\put(44,40){\footnotesize{$0.52$}}
      \put(76,38){\footnotesize{$r~(\mathrm{m})$}}
\put(61,40){\footnotesize{$0.42$}}
\put(68,40){\footnotesize{$0.44$}}
\put(74,40){\footnotesize{$0.46$}}
\put(81,40){\footnotesize{$0.48$}}
\put(88,40){\footnotesize{$0.50$}}
\put(94,40){\footnotesize{$0.52$}}
      \put(51,3){\footnotesize{$r~(\mathrm{m})$}}
\put(36,5){\footnotesize{$0.42$}}
\put(43,5){\footnotesize{$0.44$}}
\put(49,5){\footnotesize{$0.46$}}
\put(56,5){\footnotesize{$0.48$}}
\put(63,5){\footnotesize{$0.50$}}
\put(69,5){\footnotesize{$0.52$}}
        \put(13,63.5){\footnotesize{DP50}}   
        \put(13,60.5){\footnotesize{DP450}} 
        \put(63.5,63.5){\footnotesize{DP50}}   
        \put(63.5,60.5){\footnotesize{DP450}} 
        \put(38.7,28.7){\footnotesize{DP50}}   
        \put(38.7,25.7){\footnotesize{DP450}} 
      \put(0,69){\footnotesize{(a)}}
      \put(50,69){\footnotesize{(b)}}
       \put(27,34){\footnotesize{(c)}}
      \put(12.5,57){\footnotesize{$R_{I}$}}
      \put(34,57){\footnotesize{$R_{O}$}}
      \put(21,57){\footnotesize{$R_{I}$}}
      \put(39.5,57){\footnotesize{$R_{O}$}}
     \put(63,47){\footnotesize{$R_{I}$}}
      \put(84,47){\footnotesize{$R_{O}$}}
      \put(71,47){\footnotesize{$R_{I}$}}
      \put(90,47){\footnotesize{$R_{O}$}}
     \put(37,22){\footnotesize{$R_{I}$}}
      \put(59,22){\footnotesize{$R_{O}$}}
      \put(50,22){\footnotesize{$R_{I}$}}
      \put(65,22){\footnotesize{$R_{O}$}}
      \end{overpic}
\caption{ 
Shell-averaged flow field physical quantities distributed along the radial direction at time $t=2.5~\mathrm{ms}$.
The blue and red dashed lines respectively mark the characteristic radius positions of cases DP50 and DP450.
}\label{fig:profile}	
\end{figure}

To analyze the influence of particle size on flow evolution, the shell-averaged flow field quantities are calculated as
\begin{equation}
    \Bar{f}=\frac{\pi^2}{4}\int_{0}^{\pi/2}\int_{0}^{\pi/2} f~\mathrm{d}\theta \mathrm{d}\phi,~~~f=\alpha_{p},u_{fr},p_{f},
\end{equation}
where $u_{fr}$ is the gas velocity in the $r$-direction. 
Figure~\ref{fig:profile}(a) demonstrates that for cases with different particle size, $\bar{\alpha}_{p}$ near $R_{I}$ and $R_{O}$ drop below $0.1\%$ after $t=2.5~\mathrm{ms}$, indicating the particle bed transitions from an initially dense state to dilute where inter-particle collision forces become negligible.
The ratio of particle density $\rho_{p}$ to gas density $\rho_{f}$ has the magnitude of $O(10^3)$. 
Thus, the mass loading parameter $\psi_{M} = \rho_{p}\alpha_{p}/ \left(\rho_{f}\alpha_{f}\right)$ near $R_{I}$ and $R_{O}$ has the magnitude of $O(10^{-1})$ at late time, suggesting that the two-way coupling effects can be neglected~\citep{balachandar2010turbulent}.

Figure~\ref{fig:profile}(b) shows that, for both cases DP50 and DP450, a radially inward gas pressure gradient appears through the particle bed at $t=2.5~\mathrm{ms}$. 
Under the sustained effect of this radially inward pressure gradient, the gas radial velocity continues to decelerate, resulting in a lower gas radial velocity near the inner side of the particle bed compared to that near the outer side.
In case DP50, the gas pressure gradient passing through the particle bed is larger than that in case DP450. 
Consequently, as shown in Figure~\ref{fig:profile}(c), the gas radial velocity difference between positions near $R_{I}$ and $R_{O}$ is greater in case DP50 than that in case DP450.

From the equation of parcel motion in \eqref{eq:ptcdym}, the $r$-velocity $v_{pr}$ and $r$-acceleration $a_{pr}$ of parcel $i$ are contributed from the drag, pressure, and collision as~\citep{he2024duallayer}
\begin{equation}
\frac{\mathrm{D}v_{pr,i}}{\mathrm{D}t}=a_{pr,i},
\end{equation}
\begin{equation}\label{eq:apri}
	\begin{aligned}
		a_{pr, i}=\underbrace{\vphantom{\frac{1}{m_{p,i}}\sum_{j}F_{Cr,ij}}D_{p,i}\left(\boldsymbol{v}_{pr,i}-\boldsymbol{u}_{fr}\right)}_{\text{Drag},~a_{idr}}\underbrace{\vphantom{\frac{1}{m_{p,i}}\sum_{j}F_{Cr,ij}}-\frac{1}{\rho_{p}}\partial_{r} p_{f}}_{\text{Pressure},~a_{ipr}}+\underbrace{\frac{1}{m_{p,i}}\sum_{j}F_{Cr,ij}}_{\text{Collision},~a_{icr}}.
	\end{aligned}
\end{equation}
Here, $F_{Cr,ij}$ is the component of the collision force in the $r$-direction.

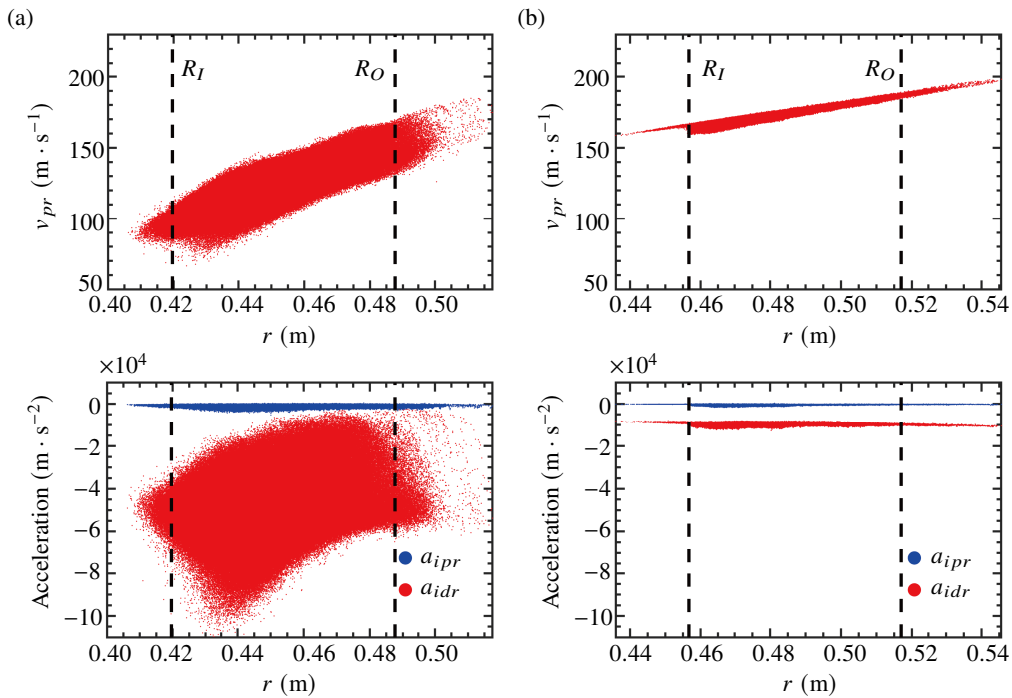
\begin{figure}
	\centering
     \begin{overpic}
	     [width=1.0\textwidth]{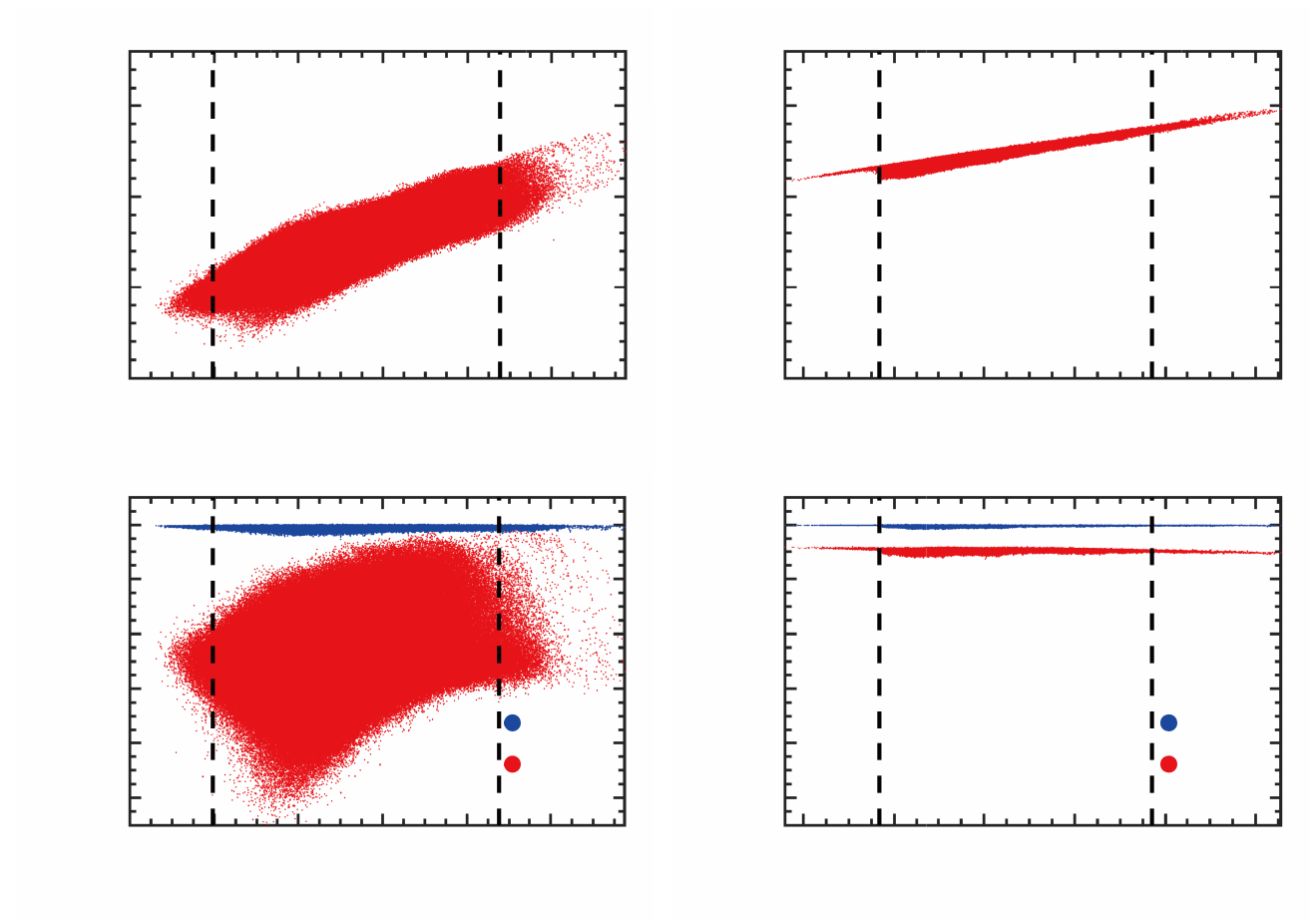}
      \put(0,67){\footnotesize{(a)}}
      \put(50,67){\footnotesize{(b)}}
      \put(17,62){\footnotesize{$R_I$}}
      \put(68,62){\footnotesize{$R_I$}}
      \put(34,62){\footnotesize{$R_O$}}
      \put(84,62){\footnotesize{$R_O$}}
\put(25,2){\footnotesize{$r~(\mathrm{m})$}}
\put(8,5){\footnotesize{$0.40$}}
\put(14,5){\footnotesize{$0.42$}}
\put(21,5){\footnotesize{$0.44$}}
\put(28,5){\footnotesize{$0.46$}}
\put(34,5){\footnotesize{$0.48$}}
\put(40,5){\footnotesize{$0.50$}}
\put(25,36){\footnotesize{$r~(\mathrm{m})$}}
\put(8,39){\footnotesize{$0.40$}}
\put(14,39){\footnotesize{$0.42$}}
\put(21,39){\footnotesize{$0.44$}}
\put(28,39){\footnotesize{$0.46$}}
\put(34,39){\footnotesize{$0.48$}}
\put(40,39){\footnotesize{$0.50$}}
\put(76,2){\footnotesize{$r~(\mathrm{m})$}}
\put(59,5){\footnotesize{$0.44$}}
\put(66,5){\footnotesize{$0.46$}}
\put(73,5){\footnotesize{$0.48$}}
\put(80,5){\footnotesize{$0.50$}}
\put(87,5){\footnotesize{$0.52$}}
\put(94,5){\footnotesize{$0.54$}}
\put(76,36){\footnotesize{$r~(\mathrm{m})$}}
\put(59,39){\footnotesize{$0.44$}}
\put(66,39){\footnotesize{$0.46$}}
\put(73,39){\footnotesize{$0.48$}}
\put(80,39){\footnotesize{$0.50$}}
\put(87,39){\footnotesize{$0.52$}}
\put(94,39){\footnotesize{$0.54$}}
\put(2,47){\rotatebox{90}{\footnotesize{$v_{pr}~(\mathrm{m\cdot s^{-1}})$}}}
\put(7,41){\footnotesize{$50$}}
\put(6,47){\footnotesize{$100$}}
\put(6,54){\footnotesize{$150$}}
\put(6,61){\footnotesize{$200$}}
\put(52,47){\rotatebox{90}{\footnotesize{$v_{pr}~(\mathrm{m\cdot s^{-1}})$}}}
\put(56.5,41){\footnotesize{$50$}}
\put(55.5,47){\footnotesize{$100$}}
\put(55.5,54){\footnotesize{$150$}}
\put(55.5,61){\footnotesize{$200$}}
\put(2,10){\rotatebox{90}{\footnotesize{$\text{Acceleration}~(\mathrm{m\cdot s^{-2}})$}}}
\put(8,29){\footnotesize{$0$}}
\put(6.5,25){\footnotesize{$-2$}}
\put(6.5,21){\footnotesize{$-4$}}
\put(6.5,17){\footnotesize{$-6$}}
\put(6.5,13){\footnotesize{$-8$}}
\put(5.5,8.5){\footnotesize{$-10$}}
\put(9,33){\footnotesize{$\times 10^{4}$}}
\put(52,10){\rotatebox{90}{\footnotesize{$\text{Acceleration}~(\mathrm{m\cdot s^{-2}})$}}}
\put(58,29){\footnotesize{$0$}}
\put(56.5,25){\footnotesize{$-2$}}
\put(56.5,21){\footnotesize{$-4$}}
\put(56.5,17){\footnotesize{$-6$}}
\put(56.5,13){\footnotesize{$-8$}}
\put(55.5,8.5){\footnotesize{$-10$}}
\put(59,33){\footnotesize{$\times 10^{4}$}}
\put(40.5,14.5){\footnotesize{$a_{ipr}$}}
\put(40.5,11.5){\footnotesize{$a_{idr}$}}
\put(90.5,14.5){\footnotesize{$a_{ipr}$}}
\put(90.5,11.5){\footnotesize{$a_{idr}$}}
\end{overpic}
\caption{%
Scatter plots of the radial velocity and radial acceleration of parcels in the cases (a)~DP50 and (b)~DP450 at $t=2.5~\mathrm{ms}$. 
The dashed lines respectively mark the characteristic radius positions.
}\label{fig:ptcDy}	
\end{figure}

The scatter plot of $v_{pr}$ and $a_{pr}$ of parcels along $r$ at $t=2.5~\mathrm{ms}$ is shown in figure~\ref{fig:ptcDy}. 
For both cases DP50 and DP450, the radial velocity of parcels decelerates such that the inner region of the particle bed is slower than the outer region. 
The radial velocity difference between the particles near $R_{I}$ and $R_{O}$ in case DP50 is around $50~\mathrm{m\cdot s^{-1}}$, which is much larger than $10~\mathrm{m\cdot s^{-1}}$ in case DP450.
In both cases, the drag coupling term $a_{dr}$ is one order of magnitude greater than the pressure gradient term $a_{pr}$.
Since that the inter-particle collision forces near $R_{I}$ and $R_{O}$ are negligible, the drag coupling term dominates the particle motion near $R_{I}$ and $R_{O}$ regions. 
Moreover, the difference in drag coupling term near $R_{I}$ and $R_{O}$ in case DP50 is significantly larger than that in case DP450.
Therefore, particles with lower inertia experience a greater difference in drag forces between the inner and outer regions of the particle bed, resulting in a larger velocity difference and ultimately accelerating the growth rate of the bed thickness.

Therefore, we demonstrate that the initial angular non-uniform distribution of particles induces an asymmetric gas radial velocity in the angular direction. 
Since the gas-particle mixture at $R_{O}$ is under a one-way coupling regime, the non-uniform gas radial velocity, coupled via gas-particle drag forces, leads to a angluar asymmetry in the radial velocity of small-inertia particles.
Consequently, well-defined particle jets are formed. 
In contrast, particles with large size, characterized by large inertial, do not exhibit significant jetting behavior.
Note that the formation of external particle jets on a cylindrical particle bed subjected to strong explosive loading is the same as described above and is elaborated in detail in Appendix~\ref{app:CY}.

Figure~\ref{fig:slice} presents local slices of the particle bed at an angular coordinate of $\theta=\pi/3$, with widths of $0.25~\mathrm{m}$ and $0.29~\mathrm{m}$ in the $z$- and $r$-directions, respectively.
In both cases, the radial gas velocity near the inner side of the particle bed is slower than that outside the bed at $t=2.5~\mathrm{ms}$, whereas it exceeds the outer-region gas radial velocity at $t=5.0~\mathrm{ms}$.
In figure~\ref{fig:slice}(a), case DP50 exhibits a small-scale particle jet structure within the bed at $t = 2.5~\mathrm{ms}$. 
By $t = 5.0~\mathrm{ms}$, the particle bed thickness shows marked growth as the particle jet evolves into a distinct, well-defined structure. 
In contrast, for case DP450 shown in figure~\ref{fig:slice}(b), the particle bed thickness increases only gradually, and no obvious jet structure is observed.

\begin{figure}
	\centering
     \begin{overpic}
	     [width=1.0\textwidth]{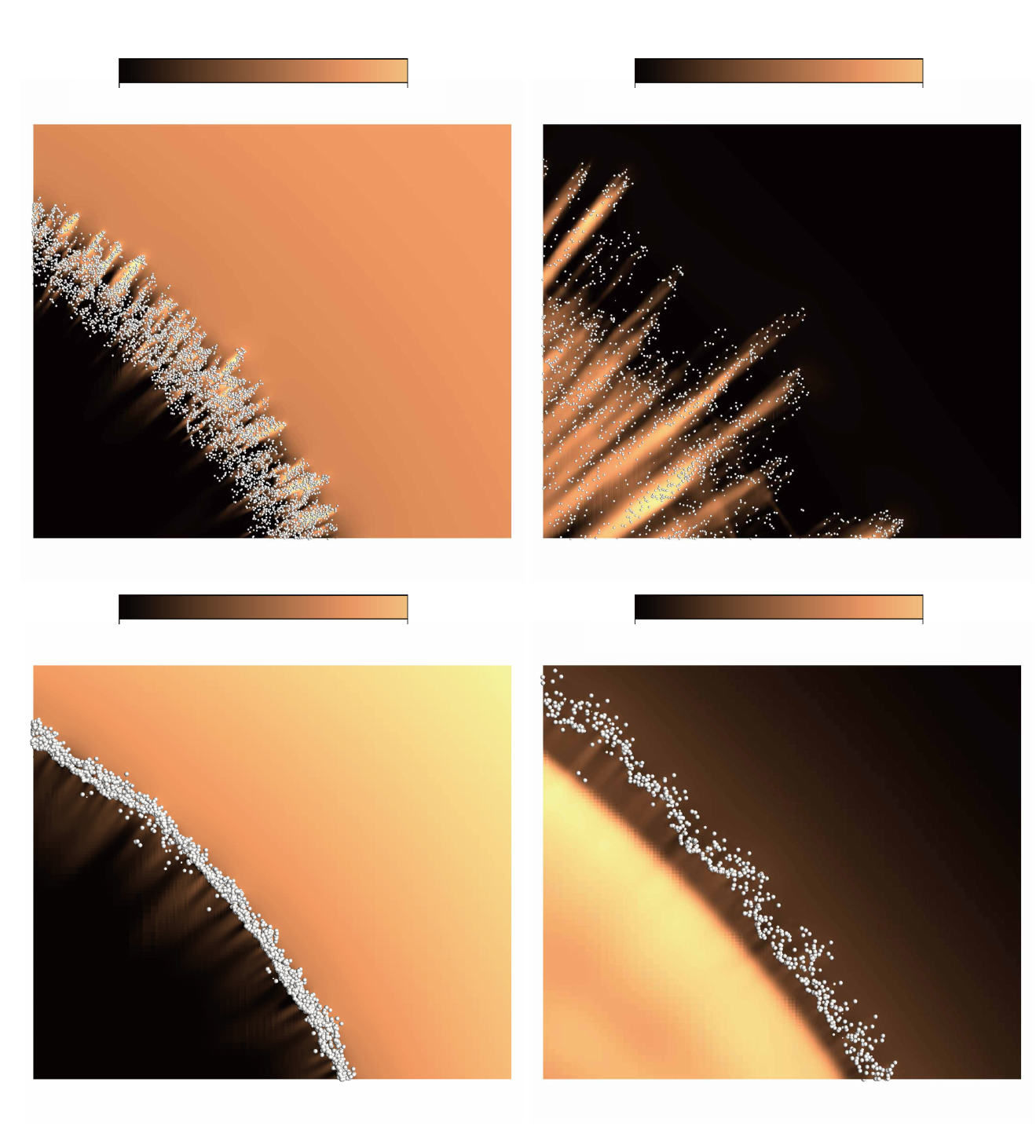}
      \put(0,97){\footnotesize{(a)}}
      \put(0,51){\footnotesize{(b)}}
      \put(18,50){\footnotesize{$u_{fr}\left(\mathrm{m\cdot s^{-1}}\right)$}}
       \put(62,50){\footnotesize{$u_{fr}\left(\mathrm{m\cdot s^{-1}}\right)$}}
       \put(18,96){\footnotesize{$u_{fr}\left(\mathrm{m\cdot s^{-1}}\right)$}}
        \put(62,96){\footnotesize{$u_{fr}\left(\mathrm{m\cdot s^{-1}}\right)$}}
        \put(8,91){\footnotesize{$-30$}}
        \put(33,91){\footnotesize{$130$}}
      \put(8,44){\footnotesize{$-30$}}
        \put(34,44){\footnotesize{$40$}}
        \put(53,91){\footnotesize{$-15$}}
        \put(78,91){\footnotesize{$120$}}
      \put(53,44){\footnotesize{$-10$}}
        \put(79,44){\footnotesize{$50$}}
        \put(18,2){\footnotesize{$t=2.5~\mathrm{ms}$}}
        \put(64,2){\footnotesize{$t=5.0~\mathrm{ms}$}}
      \end{overpic}
\caption{%
Cross-sectional views of parcel distributions and gas radial velocity fields for the cases (a)~DP50 and (b)~DP450.
}\label{fig:slice}	
\end{figure}
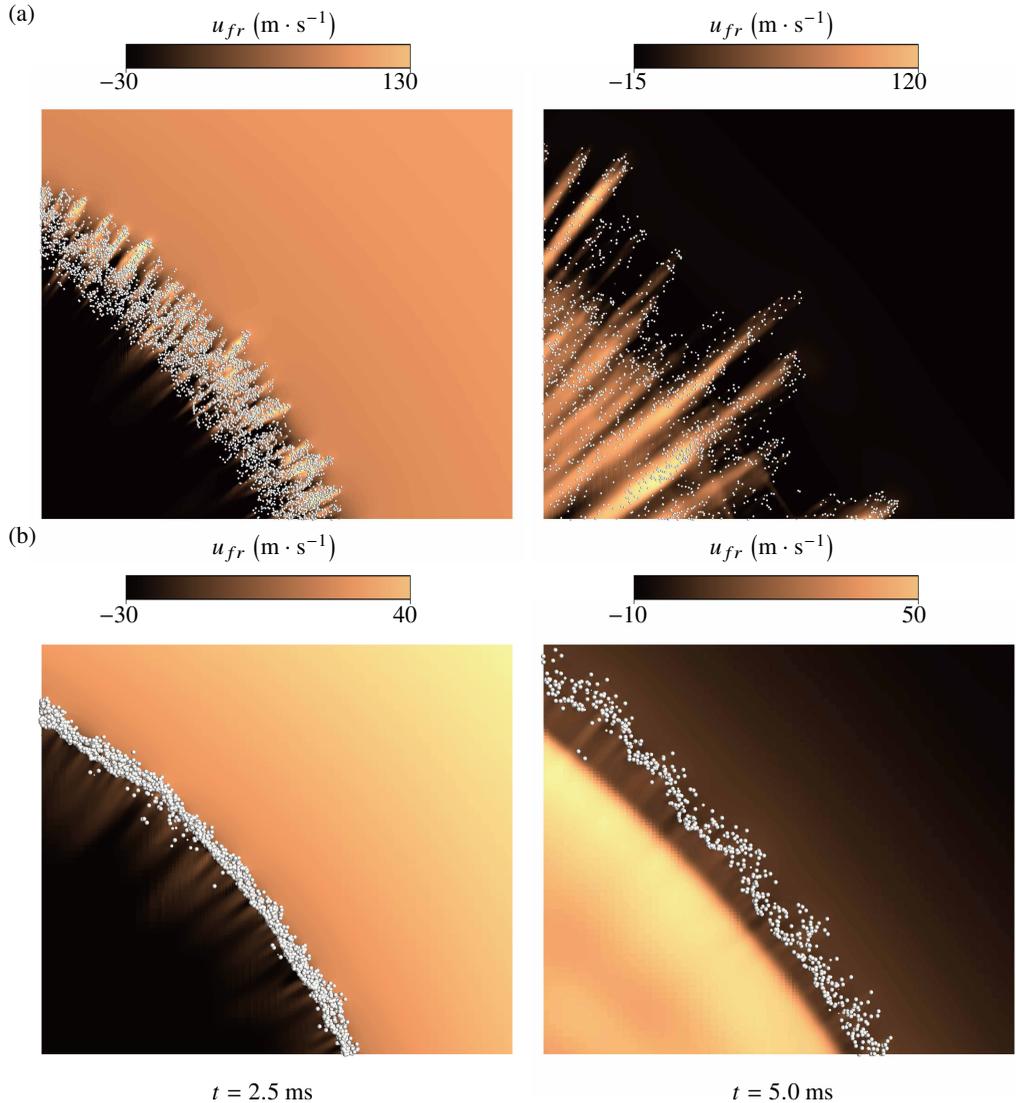

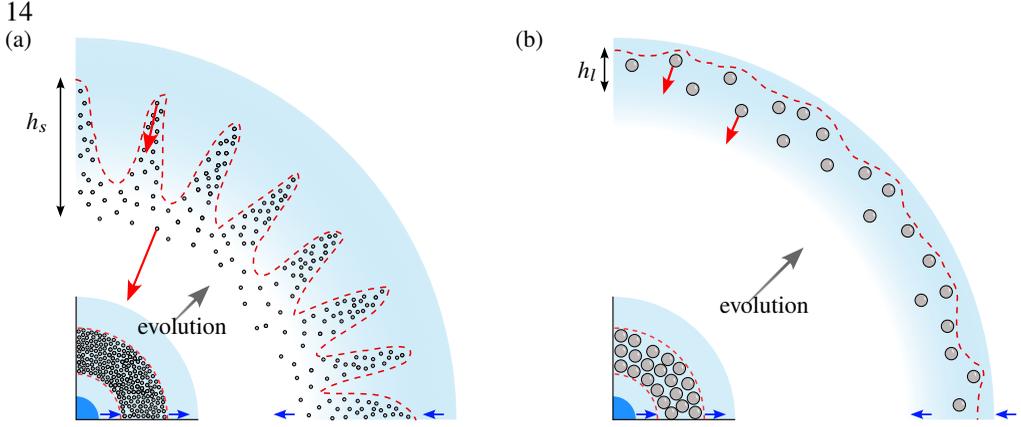
\begin{figure}
	\centering
     \begin{overpic}
	     [width=1.0\textwidth]{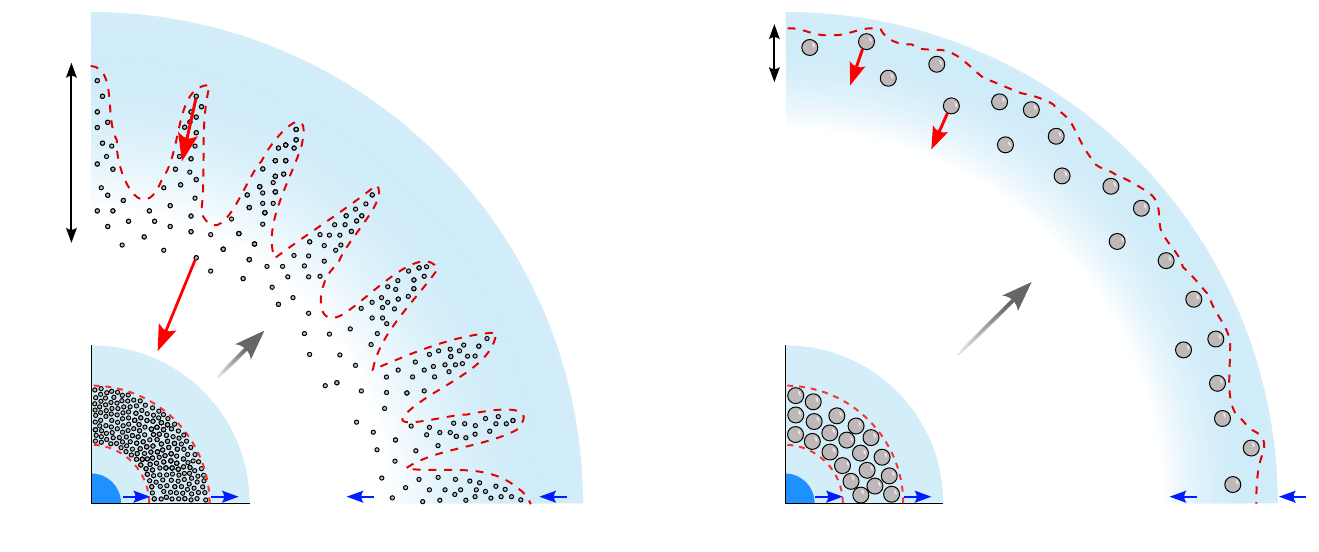}
        \put(0,40){\footnotesize{(a)}}
        \put(50,40){\footnotesize{(b)}}
        \put(2,32){\footnotesize{$h_{s}$}}
       \put(56,37){\footnotesize{$h_{l}$}}
      \put(13,12){\footnotesize{evolution}}
      \put(70,14){\footnotesize{evolution}}
	\end{overpic}
\caption{ 
Schematic for the evolution of the particle bed with (a) small $d_{p}$ and (b) large $d_{p}$.
The particle bed (gray circles) is enclosed by the red dashed line.  
The region is shaded based on pressure (darker areas indicate higher pressure). 
The red arrows represent the radial acceleration of the particles, with longer arrows denoting greater magnitudes.
The blue arrows represent the radial flow velocities. 
}\label{fig:dp_scheme}	
\end{figure}

The schematic in figure~\ref{fig:dp_scheme} illustrates the evolution of the particle bed in the spherical explosive dispersal system.
The high pressure gas pocket first accelerates the radial expansion of the particle bed, during which the particle bed transitions from a dense to a dilute state.
In the small-particle case, as shown in figure~\ref{fig:dp_scheme}(a), the particles undergo stronger radial deceleration and experience a greater acceleration difference between the inner and outer sides of the bed. 
This leads to a significant increase in $h_s$ and the emergence of distinct external particle jets on the outer side.
In contrast, for the large-particle case, depicted in figure~\ref{fig:dp_scheme}(b), the radial deceleration is weaker and the acceleration difference across the bed is smaller. 
Consequently, $h_{l}$ increases more gradually, and no pronounced jet structure is observed on the outer side.
 
In summary, the evolution of the characteristic radii $R_{I}$ and $R_O$ of the particle bed is divided into two distinct phases: a linear phase and a nonlinear deceleration phase. 
Initially, the high-pressure gas shock impact accelerates the particle bed, resulting in an approximately linear growth of both $R_{I}$ and $R_O$, where the cases with different particle sizes exhibit nearly identical expansion rates during this linear stage. 
Subsequently, as the particle bed expands, the pressure in its central region continuously decreases.
After this pressure drops below the ambient pressure, the particle bed transitions into the nonlinear deceleration phase. 
In this latter phase, the drag force dominates particle motion, leading to faster deceleration of the particle bed for the case with smaller $d_{p}$, accompanied by a greater velocity difference of particles at $R_{I}$ and $R_O$, which consequently induces the larger growth of in bed thickness.

\section{Modeling the evolution of characteristic radii in particle bed}\label{sec:model}
Now we model the evolution of the characteristic radii $R_I$ and $R_O$ of the particle bed by dividing the process into two distinct phases: the linear growth and the nonlinear deceleration growth.
Upon impingement by high-pressure gas in the central region, particles adjacent to $R_{I0}$ acquire radial velocity
\begin{equation}\label{eq:gurney}
V_{a}=F\left(\Phi_{0}, \Gamma_M\right)\sqrt{\frac{2 e_{\text {gas }}}{\frac{\Gamma_M}{0.31 \rho_{p}^{0.132}}+0.6}},
\end{equation} 
which is estimated by a modified Gurney equation~\citep{Milne2016}, with the correction function
\begin{equation}
F\left(\Phi_{0}, \Gamma_M\right)=1+\left[0.168 \exp \left(1.09 \Phi_{0}\right)-0.5\right]\log_{10}(\Gamma_M)
\end{equation}
and the energy per unit mass 
\begin{equation}
    e_{gas}=\frac{\left(P_s-P_0\right)}{\left(\gamma-1\right)\rho_s}
\end{equation}
of the gas pocket~\citep{crowl2003understanding}.
Then, $R_I$ begins to grow linearly with
\begin{equation}\label{eq:RIlinear}
R_I\left(t\right)=R_{I0}+\int_{0}^{t}V_a~\mathrm{d}t,   
\end{equation}
and the compaction front moves with velocity~\citep{xue23}
\begin{equation}\label{eq:vcf}
    V_{CF}=\sqrt{\frac{P_r-P_0}{\rho_p}\frac{\Phi_{comp}}{\left(\Phi_{comp}-\Phi_0\right)\Phi_{0}}},
\end{equation}
where the volume fraction of the 3D compacted particle bed is set to be $\Phi_{comp}=0.65$, and the pressure exerting on the inner surfaces of the particle bed is estimated by $P_r=P_{s}R^2_{g}/R^2_{I0}$.

After a duration of $t_a=h_0/V_{CF}$, the compaction wave arrives at $R_{O0}$, following which particles in the vicinity achieve velocity $V_a$ after an acceleration period of
\begin{equation}
    t_b=V_a/a_{O},
\end{equation}
where the acceleration induced by gas-particle coupling force is estimated by
\begin{equation}  a_{O}=\frac{P_sR_{g}^3/R_{I}^3\left(t_a\right)-P_0}{\rho_p h_{0}}.
\end{equation}
Then at time $t_c=t_a + t_b$, particles near $R_{O0}$ initiate the linear growth 
\begin{equation}\label{eq:ROlinear}
    R_O\left(t\right)=R_{O0}+\int_{t_a}^{t_c}\int_{t_a}^{t_c}a_{O}~\mathrm{d}t\mathrm{d}t+\int_{t_c}^{t}V_a~\mathrm{d}t.
\end{equation}

When the particle bed expands to $t=t_d$, the central pressure equilibrates with ambient pressure.
Neglecting the gas leakage and assuming that the gas in the central zone undergoes isothermal expansion, we obtain
\begin{equation}
    \frac{R_{O}^3\left(t_{d}\right)}{R_{g}^3}\approx\frac{P_{s}}{P_{0}},
\end{equation} 
at which point the characteristic radius transitions into a nonlinear deceleration phase,
\begin{equation}\label{eq:nonlinear}    R_{\Psi}\left(t\right)=R_{\Psi}\left(t_{d}\right)+\int_{t_{d}}^{t}\left(V_{a}+\int_{t_d}^{t}a_{dr,\Psi}~\mathrm{d}t\right)\mathrm{d}t,~~~\Psi=O,~I,
\end{equation}
where the drag coupling term $a_{dr,\Psi}$ dominates the motion of particles near $R_{I}$ and $R_{O}$.
Combining the dilute flow regime $\alpha_{f}\approx1$ and the drag model in \eqref{eq:drag} and \eqref{eq:cd}, $a_{dr,\Psi}$ can be expressed as
\begin{equation}\label{eq:adr}
    a_{dr,\Psi}\left(t\right)\approx-\frac{18\rho_{f}}{f_{base}\mathrm{Re}_{p}\rho_{p}d_{p}}\left(v_{pr}-u_{fr}\right)^2,
\end{equation}
where $f_{base}$ is defined in~\eqref{eq:fbase}, and particle radial velocity dominated by drag is approximated as
\begin{equation}\label{eq:ufr}
v_{pr}\approx V_{a}+\int_{t_d}^{t}a_{dr,\Psi}~\mathrm{d}t.
\end{equation}

To close the model of nonlinear deceleration phase in~\eqref{eq:gurney},~\eqref{eq:nonlinear}, ~\eqref{eq:adr} and~\eqref{eq:ufr}, we introduce an empirical parameter $\beta_{\Psi}$ to model the velocity difference between particle and gas
\begin{equation}\label{eq:beta}
|v_{pr}-u_{fr}|=\beta_{\Psi} v_{pr},~~~\Psi=O, I.
\end{equation}
During the deceleration phase, the gas radial velocity $u_{fr}$ and particle radial velocity $v_{pr}$ are in opposite directions in the vicinity of $R_I$, whereas they are aligned near $R_{O}$~(shown in figures~\ref{fig:profile} and~\ref{fig:ptcDy}). 
Consequently, we expect $\beta_I>1$ and $\beta_O<1$. 

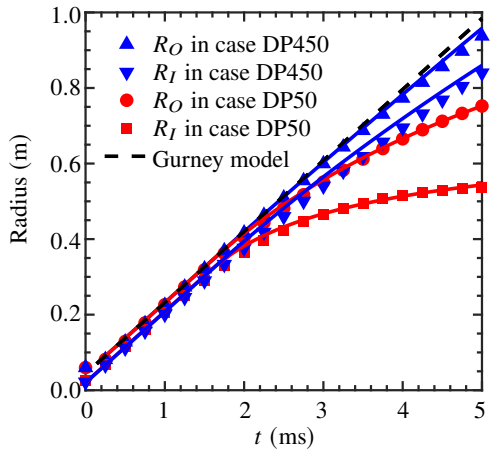
\begin{figure}
	\centering
     \begin{overpic}
	     [width=0.5\textwidth]{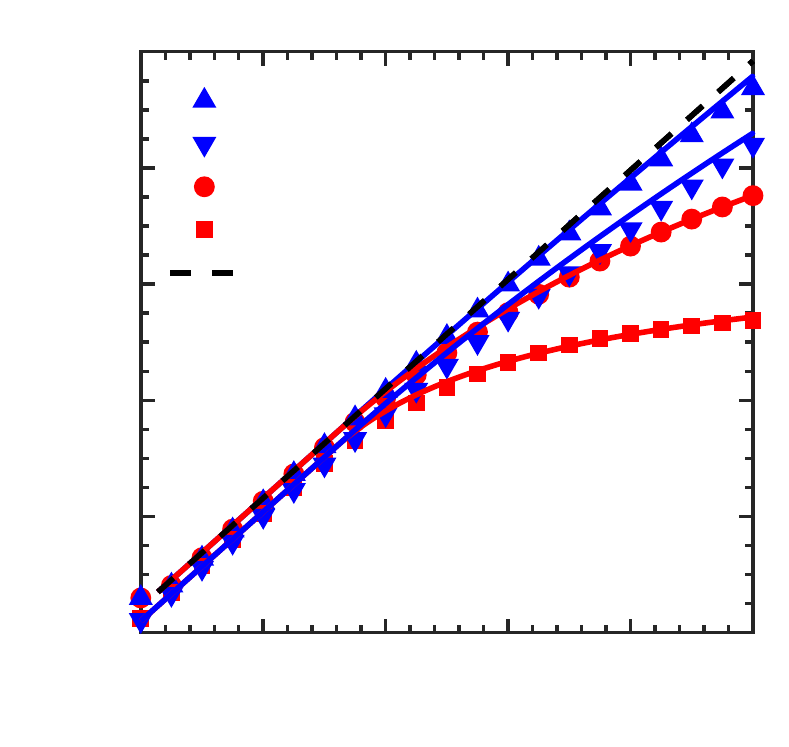}
        \put(31,80.5){\footnotesize{$R_{O}$ in case DP450}}   
        \put(31,75){\footnotesize{$R_{I}$ in case DP450}} 
        \put(31,69.5){\footnotesize{$R_{O}$ in case DP50}}   
        \put(31,64){\footnotesize{$R_{I}$ in case DP50}} 
        \put(31,58){\footnotesize{Gurney model}} 
        \put(11,13){\footnotesize{$0.0$}}
        \put(11,28){\footnotesize{$0.2$}}
        \put(11,42){\footnotesize{$0.4$}}
        \put(11,57){\footnotesize{$0.6$}}
        \put(11,72){\footnotesize{$0.8$}}
        \put(11,86){\footnotesize{$1.0$}}
        \put(17,9){\footnotesize{$0$}}
        \put(32,9){\footnotesize{$1$}}
        \put(48,9){\footnotesize{$2$}}
        \put(63,9){\footnotesize{$3$}}
        \put(79,9){\footnotesize{$4$}}
        \put(94,9){\footnotesize{$5$}}
        \put(51,4){\footnotesize{$t~(\mathrm{ms})$}}
       \put(3,43){\rotatebox{90}{\footnotesize{$\text{Radius}~(\mathrm{m})$}}}
	\end{overpic}
\caption{ 
Evolution of characteristic radii from the Gurney model (dashed lines), numerical simulations (symbols), and current modelling (lines).
}\label{fig:model}	
\end{figure}

Figure \ref{fig:model} shows that the model results in~\eqref{eq:RIlinear},~\eqref{eq:ROlinear} and~\eqref{eq:nonlinear} generally agree with the simulation results.
Note that parameters $\beta_I=1.5$ and $\beta_O=0.7$ are fitted from the simulation data.
For both cases DP50 and DP450, $\Gamma_{M}=6.8$ is given by~\eqref{eq:mass_ratio}. 
The Gurney model demonstrates adequate predictive capability for the evolution of characteristic radius when dealing with large particle sizes, whereas significant deviations emerge as particle size decreases. 
The current model incorporates the drag-governed nonlinear deceleration phase, characterized by \eqref{eq:ufr}, into the Gurney model.
This enhanced formulation achieves accurate prediction of characteristic radius evolution, thereby enabling capture of the primary structural features within particle beds.

Based on particle dynamics analysis, the Gurney model in \eqref{eq:gurney}~\citep{Milne2016} has been extended into the nonlinear regime by accounting for the drag-dominated deceleration phase. 
By introducing a single empirical coefficient, $\beta_{\Phi}$, the velocity difference between particles and gas is simplified in the modelling.
This leads to a concise expression for the evolution of characteristic radii $R_{\Phi}$ in~\eqref{eq:nonlinear}, facilitating straightforward comparisons with future results.

\section{Conclusions}\label{sec:conclusion}
The formation of external particle jets on a 3D spherical particle bed under strong explosive loading is examined through large-scale numerical simulations. 
An Eulerian-Lagrangian approach implemented on an adaptive mesh is utilized, with the analysis focusing on two cases of differing particle sizes.
In the large-particle case with $d_{p}=450~\mathrm{\upmu m}$, the computational mesh has an effective resolution equivalent to $1024^{3}$ grid points, and approximately $0.54$ million parcels are tracked. 
In the small-particle case with $d_{p}=50~\mathrm{\upmu m}$, a higher effective resolution of $2048^{3}$ is used, and about $1.8$ million parcels are tracked in total. 

The numerical results indicate that particle size significantly influences the evolution of the particle-bed structure. 
Pronounced particle jets are observed in the small-particle case, whereas no distinct jets develops in the large-particle case, which is consistent with experimental findings. 
Furthermore, the thickness of the particle bed in the small-particle case exhibits faster growth than that in the large-particle case. 
To quantitatively characterize the particle bed evolution, characteristic radii are defined as the inner radius $R_I$ and outer radius $R_O$. 
It is found that in both small- and large-particle cases, the characteristic radii undergoes a phase of linear growth followed by a nonlinear decelerating growth. 

While weak shock waves drive particle jets through strong vorticity-induced clustering, we find that under strong explosive loading jet formation exhibits only a weak correlation with angular clustering.
An analysis of the flow field and particle dynamics reveals that the weak correlation is attributed to the dominance of the drag force, which governs particle motion and jet formation during the nonlinear deceleration growth stages of the characteristic radii.

In the nonlinear decelerating growth stage, the flow is in a dilute state, and the effects of inter-particle collision forces are negligible. 
Meanwhile, for both the small- and large-particle cases, the drag coupling term $R_I$ and $R_O$ is significantly greater than the pressure gradient term.
Then the drag force dominates particle motion, leading to faster deceleration of the particles in the case with smaller $d_{p}$, accompanied by a greater velocity difference of particles at $R_{I}$ and $R_O$, which consequently induces the larger growth of the particle bed thickness.
The non-uniform gas radial velocity, induced by initial angular non-uniform distribution of particles, subsequently leads to an angular asymmetry in the radial velocity of small-inertia particles via gas-particle drag coupling, consequently promoting the formation of pronounced particle jets. 
In contrast, particles with large size, due to their large inertia, do not exhibit significant jetting behavior.

Finally, we establish a model with a single empirical parameter for the evolution of characteristic radii, dividing the process into linear growth and nonlinear decelerating growth phases. 
For the linear growth phase, the Gurney model is adopted. 
In the nonlinear phase, based on our prior analysis, the deceleration induced by drag force is incorporated. 
This model captures the evolution of $R_I$ and $R_O$ under different particle sizes, demonstrating good agreements with numerical simulations.

Beyond particle size, the formation of external particle jets in spherical explosive dispersal systems is governed by multiple factors.
For instance, particle fragmentation~\citep{marr2018suppression} and liquid saturation (e.g., water infiltration)~\citep{Loiseau2018} can significantly alter the dispersal pattern.
These factors necessitate more detailed physical modelling and should be incorporated into future numerical simulation studies.

\backsection[Acknowledgements]{Numerical simulations were carried out on the TianheXY-C supercomputer in Guangzhou, China.
The authors thank D. Frost for sharing the experimental images.
}

\backsection[Funding]{This work has been supported in part by the National Natural Science Foundation of China (grant
nos 12588201, 12525201, 12432012, 12432010 and 12202072).}

\backsection[Declaration of interests]{The authors report no conflict of interest.}

\backsection[Author contributions]{
Y.Y., Y.H. and B.T. designed research. 
J.Z., Y.H. and B.T. developed the simulation code. 
Y.H. conducted simulations. Y. H. and Y. Y. analyzed simulation data. 
All the authors discussed the results and wrote the manuscript. All the authors have given approval for the manuscript.
}

\appendix
\section{Comparative analysis with experimental results}\label{app:cpwithex}
In the present study, the particle properties, the geometric dimensions of the particle bed, and the parameters of the central high-pressure gas packet in the numerical cases were selected with reference to the experiments conducted by \citet{marr2018suppression}. 
Since SiC particles were used in the experiments by~\citet{marr2018suppression}, the particle density in the numerical cases was correspondingly set to $\rho_{p}=3200~\mathrm{kg\cdot m^{-3}}$.
The diameter of the particle bed in the experiments was $121~\mathrm{mm}$, corresponding to $R_{O0} = 60~\mathrm{mm}$ in the numerical cases. 
A spherical C4 explosive charge with a mass of $28~\mathrm{g}$, yielding a heat of explosion of $3.3\times10^{4} ~\mathrm{cal}$~\citep{koch2021high}, was employed in the experiments. 
Correspondingly, the internal energy of the high-pressure gas packet in the numerical case was $1.0\times10^{4}~\mathrm{cal}$, placing both internal energy values within the same order of magnitude.

\begin{figure}
	\centering
     \begin{overpic}
	     [width=0.5\textwidth]{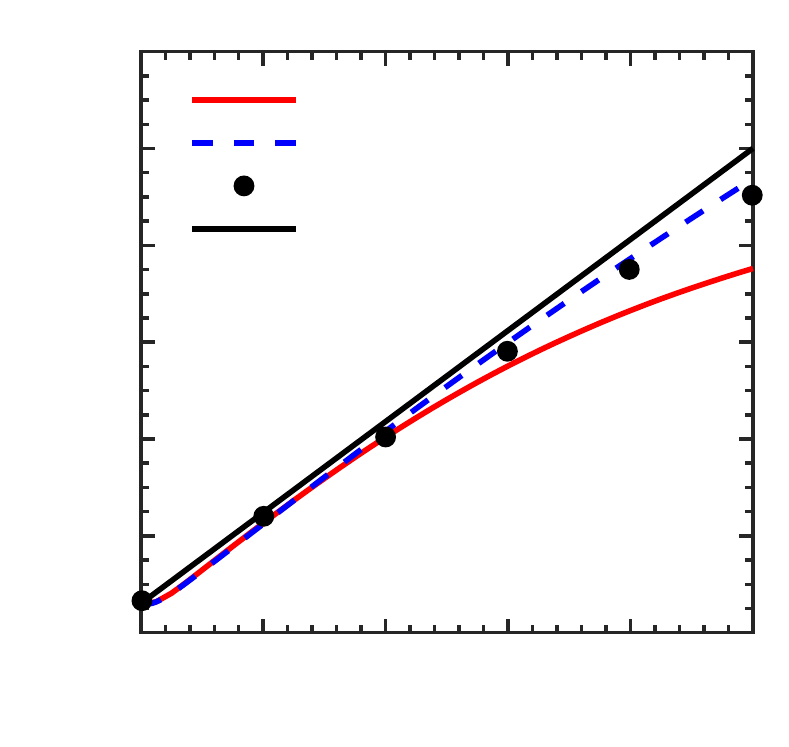}
        \put(39,81){\footnotesize{DP50}}   
        \put(39,75.5){\footnotesize{DP450}} 
        \put(39,70){\footnotesize{\citet{marr2018suppression}}} 
        \put(39,64.5){\footnotesize{Gurney model}} 
        \put(11,13){\footnotesize{$0.0$}}
        \put(11,25){\footnotesize{$0.2$}}
        \put(11,37){\footnotesize{$0.4$}}
        \put(11,49){\footnotesize{$0.6$}}
        \put(11,62){\footnotesize{$0.8$}}
        \put(11,74){\footnotesize{$1.0$}}
        \put(11,86){\footnotesize{$1.2$}}
        \put(17,9){\footnotesize{$0$}}
        \put(32,9){\footnotesize{$1$}}
        \put(48,9){\footnotesize{$2$}}
        \put(63,9){\footnotesize{$3$}}
        \put(79,9){\footnotesize{$4$}}
        \put(94,9){\footnotesize{$5$}}
        \put(51,4){\footnotesize{$t~(\mathrm{ms})$}}
       \put(3,43){\rotatebox{90}{\footnotesize{$R_{O}~(\mathrm{m}$)}}}
	\end{overpic}
\caption{%
Evolution of characteristic radius $R_O$ in the Gurney model, numerical simulations and experiments~\citep{marr2018suppression}.
}\label{fig:cpEX}	
\end{figure}
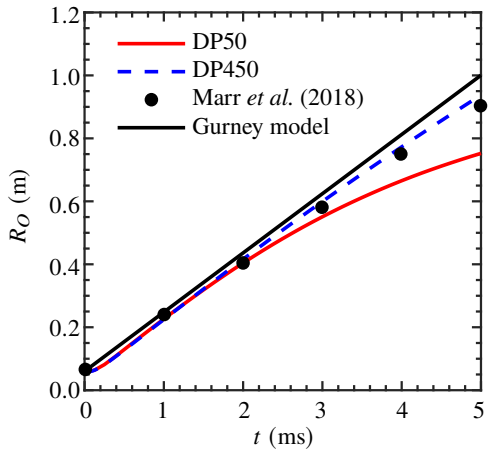

Figure~\ref{fig:cpEX} demonstrates that the evolution of $R_O$ in the numerical case DP450 agrees well with Marr's experimental results for SiC particle radius $d_{p}=430~\mathrm{\upmu m}$.
Notably, the energy of the high-pressure gas pocket in the simulation is approximately three times lower than the energy of experimental C4. 
This setting aligns with the experimental $R_O$ being two to three times lower than the Gurney model predictions using C4 explosive energy as input, indicating unaccounted energy losses (e.g., particle compaction and fragmentation) in the model. 
When substituting the simulated gas pocket energy into the Gurney model in~\eqref{eq:gurney}, the results show good agreement with experiments.

\section{Angular clustering of particles}\label{app:angular}
Under weak explosion conditions, the sustained lateral granular flow emerging from the interaction between a weak shock wave and a granular bed acts as a prerequisite for initiating particle jet growth~\citep{li2022shock}. 
Consequently, the angular position of non-uniform particle clustering under weak explosion conditions is strongly correlated with the angular position of the particle jet.
Here, under strong explosion conditions, the relationship between particle jet formation and angular clustering of particles needs to be clarified.

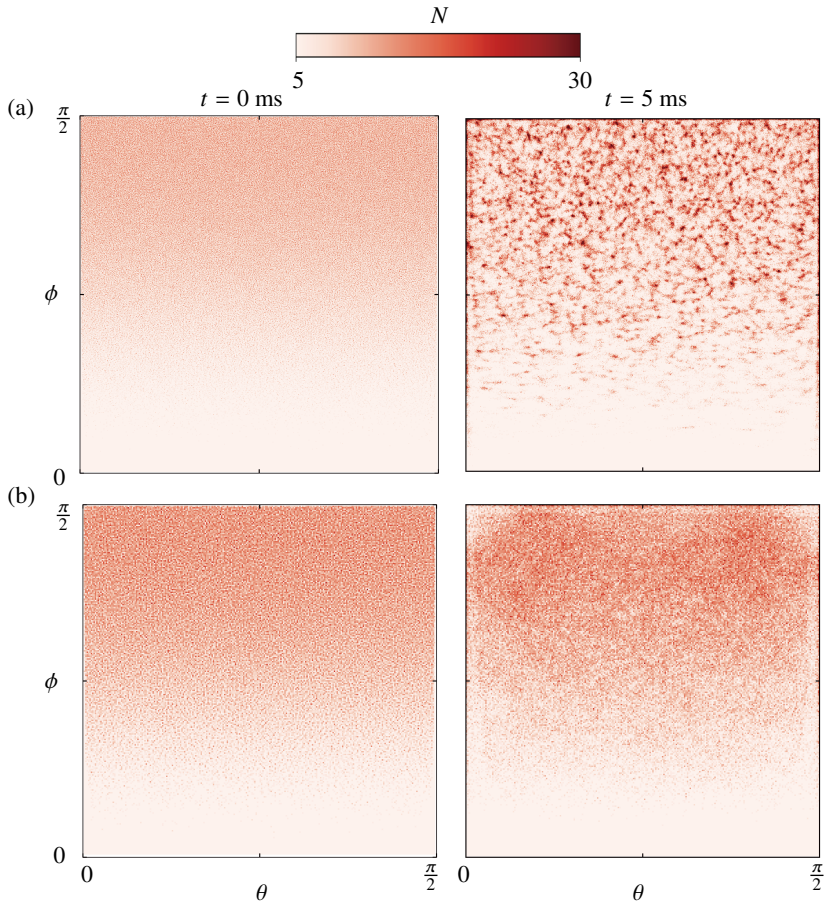
\begin{figure}
	\centering
     \begin{overpic}
	     [width=0.8\textwidth]{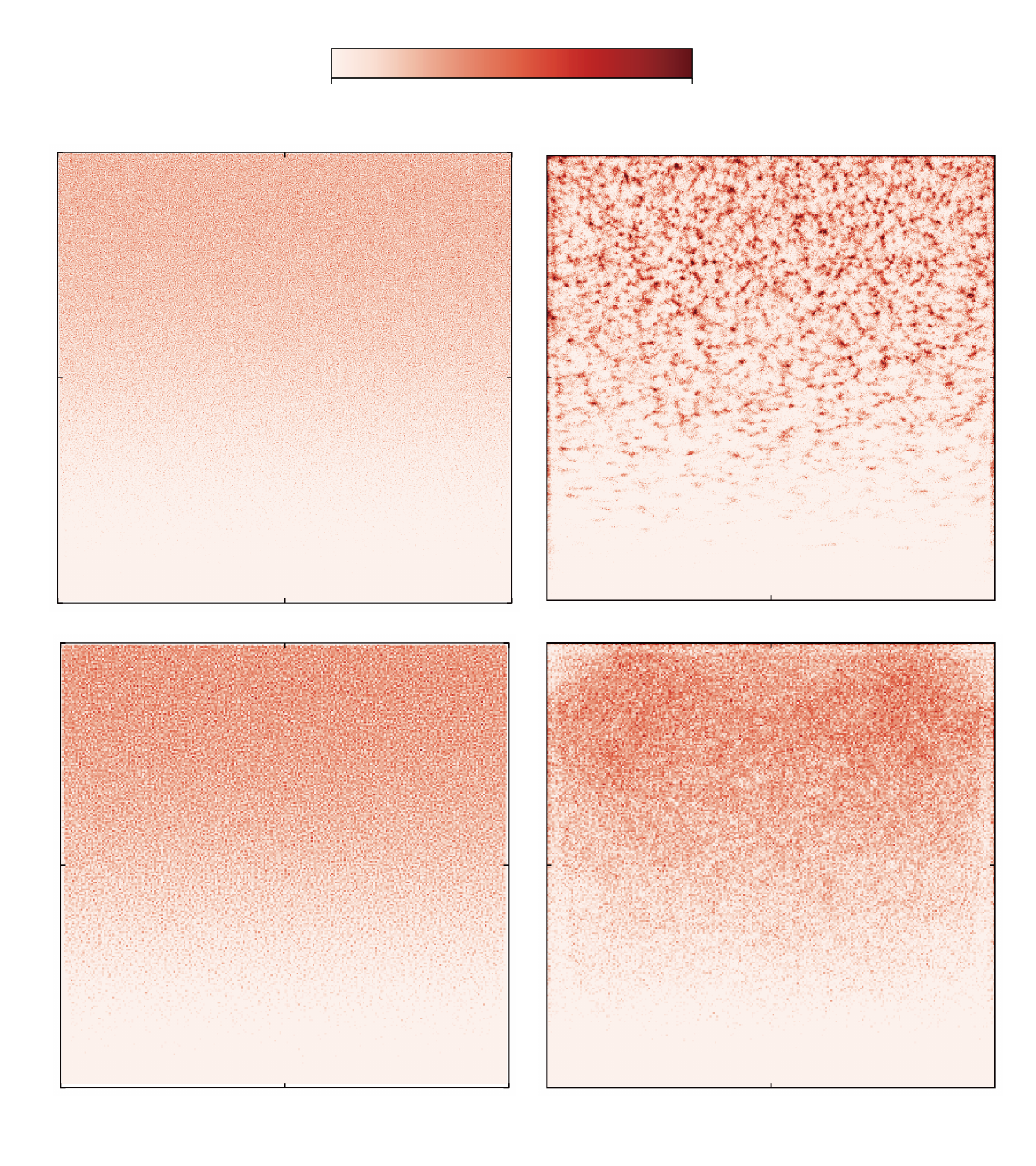}
      \put(-3,87){\footnotesize{(a)}}
      \put(-3,45){\footnotesize{(b)}}
      \put(43,97){\footnotesize{$N$}}
      \put(28,90){\footnotesize{$5$}}
      \put(58,90){\footnotesize{$30$}}
     \put(18,88){\footnotesize{$t=0~\mathrm{ms}$}}
      \put(62,88){\footnotesize{$t=5~\mathrm{ms}$}}
      \put(5,4){\footnotesize{$0$}}
      \put(42,4){\footnotesize{$\frac{\pi}{2}$}}
     \put(46,4){\footnotesize{$0$}}
      \put(84,4){\footnotesize{$\frac{\pi}{2}$}}
      \put(2,6){\footnotesize{$0$}}
      \put(2,43){\footnotesize{$\frac{\pi}{2}$}}
     \put(2,47){\footnotesize{$0$}}
      \put(2,86){\footnotesize{$\frac{\pi}{2}$}}
      \put(1,67){\footnotesize{$\phi$}}
      \put(1,25){\footnotesize{$\phi$}}
       \put(24,2){\footnotesize{$\theta$}}
      \put(65,2){\footnotesize{$\theta$}}
      \end{overpic}
\caption{ 
Evolution of total parcel number ($N$) per counting box on the angular ($\theta$-$\phi$) plane for cases (a) DP50 and (b) DP450.
}\label{fig:ang1}	
\end{figure}

Figure~\ref{fig:ang1} depicts the particle distribution on the angular ($\theta$-$\phi$) plane.
Referring to the box-counting method~\citep{Wang1993,Fessler1994},   we partition the angular plane in cases DP50 and DP450 into counting boxes with grid sizes $\pi / 1024$ and $\pi / 512$, respectively, and then count the total parcel number $N$ within each counting box.
Figure~\ref{fig:ang1} indicates that initially, both cases exhibit a uniform angular distribution of particles.
However, at $t = 5~\mathrm{ms}$, significant non-uniform angular clustering of particles emerges in case DP50, whereas no pronounced aggregation is observed in case DP450.

\begin{figure}
	\centering
     \begin{overpic}
	     [width=1.0\textwidth]{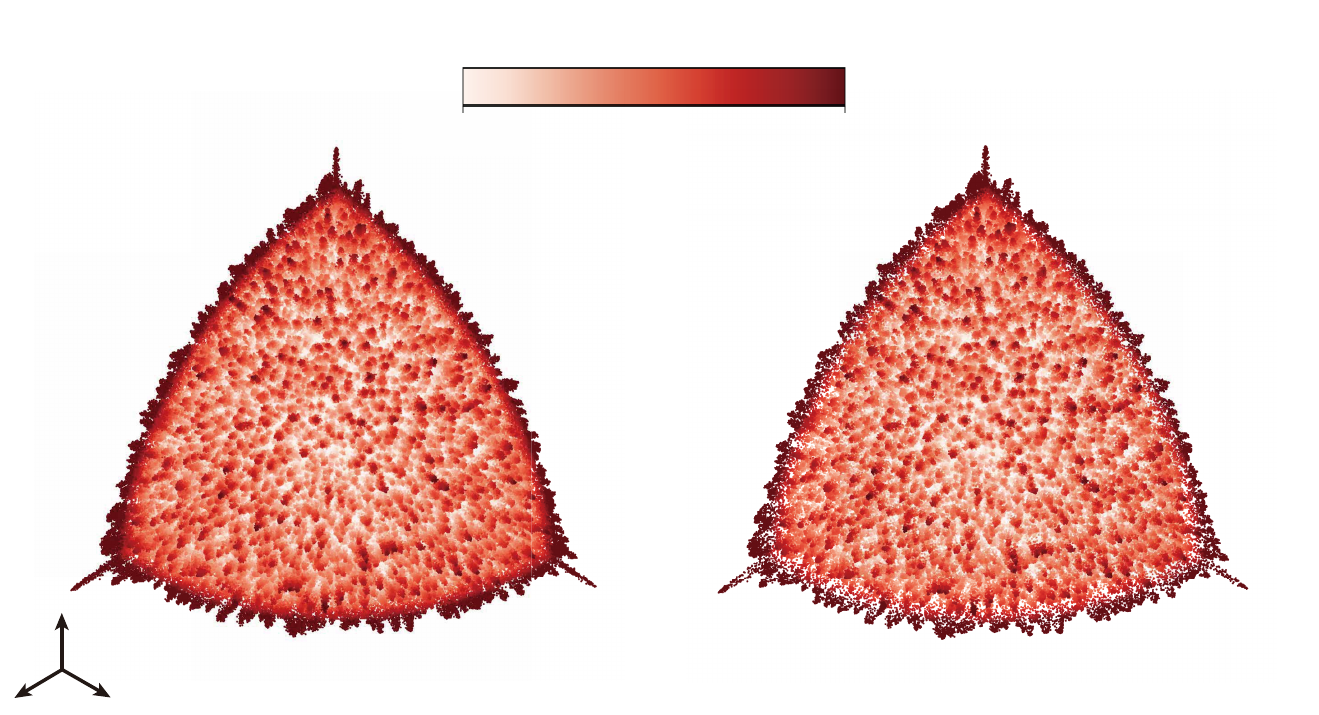}
      \put(3,40){\footnotesize{(a)}}
      \put(53,40){\footnotesize{(b)}}
      \put(46,50){\footnotesize{$r~
      (\mathrm{m})$}}
      \put(33,43){\footnotesize{$0.65$}}
      \put(61,43){\footnotesize{$0.75$}}
      \put(1.5,3){\footnotesize{$x$}}
      \put(7,3){\footnotesize{$y$}}
     \put(4,8){\footnotesize{$z$}}
      \end{overpic}
\caption{ 
Parcel distributions for case DP50 at $t=5~\mathrm{ms}$: (a) all parcels and (b) parcels with radial position $r$ in the top $10\%$ within each counting box on the angular ($\theta$-$\phi$) plane (hereinafter referred to as ``top $10\%$ particles'').
}\label{fig:ang2}	
\end{figure}

To characterize the location of the particle jet in case DP50, particles within the top $10\%$ of radial position $r$ in each counting box on the angular plane (hereinafter referred to as ``top $10\%$ particles'') are extracted.
Note that if the total parcel number in a counting box was less than $10$, the single parcel with the largest radial coordinate was selected instead.
Figure~\ref{fig:ang2} demonstrate that the top $10\%$ particles within each counting box on the angular plane can effectively characterize the jet.

\begin{figure}
	\centering
     \begin{overpic}
	     [width=1.0\textwidth]{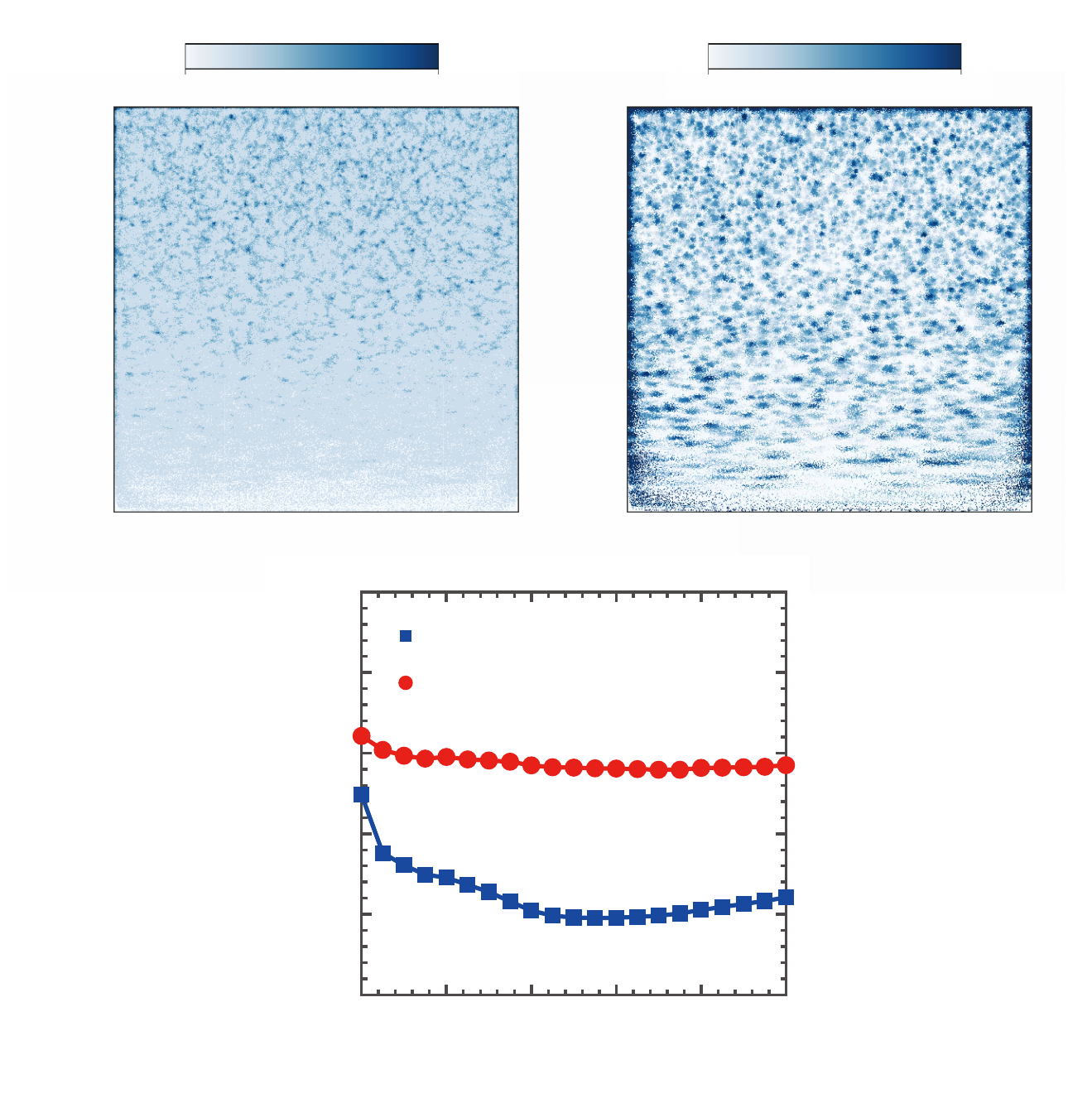}
      \put(2,97){\footnotesize{(a)}}
      \put(23,46){\footnotesize{(b)}}
     \put(8,53){\footnotesize{$0$}}
      \put(8,90){\footnotesize{$\frac{\pi}{2}$}}
      \put(54,53){\footnotesize{$0$}}
      \put(54,90){\footnotesize{$\frac{\pi}{2}$}}
     \put(10,51){\footnotesize{$0$}}
      \put(46,51){\footnotesize{$\frac{\pi}{2}$}}
      \put(56,51){\footnotesize{$0$}}
      \put(92,51){\footnotesize{$\frac{\pi}{2}$}}
      \put(5,71){\footnotesize{$\phi$}}
      \put(51,71){\footnotesize{$\phi$}}
       \put(28,49){\footnotesize{$\theta$}}
      \put(75,49){\footnotesize{$\theta$}}
      \put(26,97){\footnotesize{$N_{10\%}$}}
      \put(16,92){\footnotesize{$0$}}
      \put(39,92){\footnotesize{$5$}}
      \put(71,97){\footnotesize{$R_{10\%}~(\mathrm{m})$}}
      \put(62,92){\footnotesize{$0.65$}}
      \put(85,92){\footnotesize{$0.75$}}
      \put(26,22){\rotatebox{90}{\footnotesize{Correlation}}}
      \put(29,9){\footnotesize{$0.3$}}
      \put(29,16.5){\footnotesize{$0.4$}}
     \put(29,24){\footnotesize{$0.5$}}
      \put(29,31.5){\footnotesize{$0.6$}}
      \put(29,39){\footnotesize{$0.7$}}
      \put(29,46){\footnotesize{$0.8$}}
      \put(49,5){\footnotesize{$t~(\mathrm{ms})$}}
      \put(32,8){\footnotesize{$0$}}
     \put(39.5,8){\footnotesize{$1$}}
      \put(47.5,8){\footnotesize{$2$}}
      \put(55,8){\footnotesize{$3$}}
      \put(63,8){\footnotesize{$4$}}
      \put(70,8){\footnotesize{$5$}}
      \put(38,42){\footnotesize{$\rho\left[ N, R_{10\%}\right]$}}
      \put(38,38){\footnotesize{$\rho\left[ N_{10\%}, R_{10\%}\right]$}}
      \end{overpic}
\caption{ 
(a)~For case DP50 at $t=5~\mathrm{ms}$, total parcel number $N_{10\%}$ and mean radial coordinate $R_{10\%}$ of top $10\%$ particles.
(b)~Evolution of Pearson's correlation coefficient between $R_{10\%}$ and $N$ , and between $R_{10\%}$ and $N_{10\%}$ in case DP50.
}\label{fig:ang3}	
\end{figure}

For case DP50, we count the total parcel number of top $10\%$ particles in each counting box, denoted as $N_{10\%}$, and compute the average radial position of these particles, denoted as $R_{10\%}$.
The two parameters characterize different aspects: $N_{10\%}$ characterizes the location of angular non-uniform particle clustering, with a larger value corresponding to stronger local aggregation, and $R_{10\%}$ characterizes the angular location of the particle jet, with a larger value indicating a more pronounced jet.
As shown in figure~\ref{fig:ang3}(a), the significant non-uniform distributions of $N_{10\%}$ and $R_{10\%}$ on the angular plane differ markedly in morphology—$N_{10\%}$  being filamentary and $R_{10\%}$ being spot-like.

To quantify the discrepancy between the particle jet location and the angular non-uniform clustering of particles, we calculate the Pearson correlation coefficients~\citep{Dacome2025} between $R_{10\%}$ and $N$, and between $R_{10\%}$ and $N_{10\%}$, denoted as $\rho\left[ N, R_{10\%}\right]$ and $\rho\left[ N_{10\%}, R_{10\%}\right]$, respectively.
Figure~\ref{fig:ang3} indicates that both $\rho\left[ N, R_{10\%}\right]$ and $\rho\left[ N_{10\%}, R_{10\%}\right]$ exhibit a decreasing trend over time, although a slight increase in $\rho\left[ N, R_{10\%}\right]$  is observed after $t > 3~\mathrm{ms}$. 
Consequently, under strong explosion conditions, the formation of external particle jets is weakly correlated with the angular clustering of particles.
This stands in sharp contrast to the scenario under weak explosive shock, where vorticity deposition—induced by RMI—channels the particles into prominent, well-defined jets through angular clustering~\citep{koneru2020numerical}.

\section{Simulation results for the two-dimensional cylindrical case}\label{app:CY}
We consider a two-dimensional (2D) cylindrical configuration with the same dimensionless mass parameter $\Gamma_M = 0.68$ as the 3D simulation case.
The initial inner and outer radii of the particle bed are set to $R_{I0} = 20~\mathrm{mm}$ and $R_{O0} = 60~\mathrm{mm}$, respectively.
The annular particle bed with an initial volume fraction of $\Phi_0 = 0.50$ is filled with randomly distributed computational parcels \citep{Lozano16}.
Approximately $40,000$ parcels are employed in the simulation.%
The particles possess physical properties identical to those in case DP50.
The radius, pressure, and temperature of the gas pocket are set to $R_{g} = 16~\mathrm{mm}$, $P_{s} = 2461.5~\mathrm{bar}$, and $T_{s} = 299.5~\mathrm{K}$, respectively.
Note that in the cylindrical configuration, the mass ratio between the surrounding particles and the gas pocket is estimated as
\begin{equation}
\Gamma_M =\frac{(R_{O0}^2-R_{I0}^2) \Phi_0 \rho_p}{R_{g}^2 \rho_{s}}.
\end{equation}

The cylindrical configuration is simulated in the 2D domain $\mathcal{D} = [0, W_{x}] \times [0, W_{y}]$, with domain sizes $W_x=W_y=4~\mathrm{m}$.
The computational domain is discretized using a base grid resolution of $2048^2$, which is refined by a factor of two across two levels, resulting in a finest grid resolution of $8192^2$ and a grid size of $\Delta x = 0.5~\mathrm{mm}$.
Outflow boundary conditions are applied in both the $x$- and $y$-directions.
Both the gas pocket and the particle bed are centered at $(x_c, y_c) = (W_{x}/2, W_{y}/2)$.

\begin{figure}
	\centering
     \begin{overpic}
	     [width=1.0\textwidth]{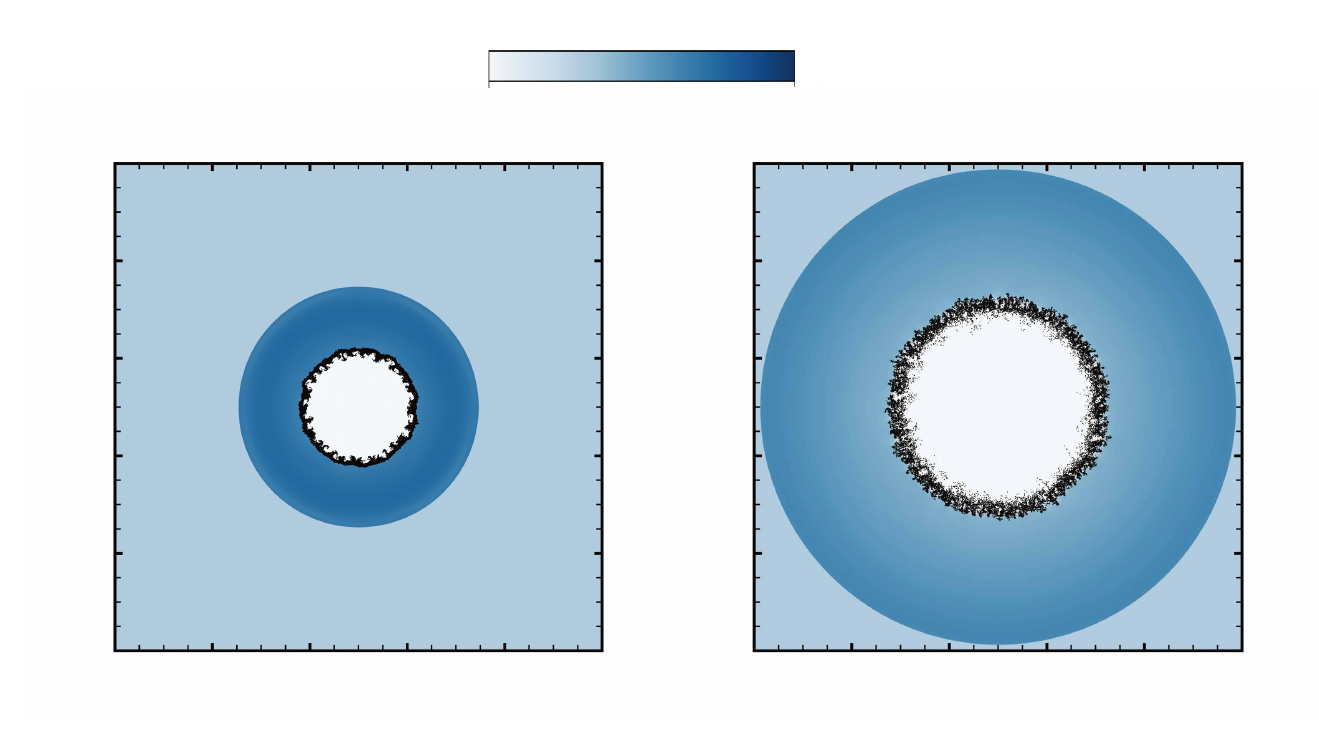}
      \put(0,46){\footnotesize{(a)}}
      \put(51,46){\footnotesize{(b)}}
      \put(45,53.5){\footnotesize{$p_f~(\mathrm{bar})$}}
      \put(36.5,47.5){\footnotesize{$0.7$}}
       \put(58,47.5){\footnotesize{$1.4$}}
      \put(25,2){\footnotesize{$x~(\mathrm{m})$}}
       \put(74,2){\footnotesize{$x~(\mathrm{m})$}}
      \put(2,24){\rotatebox{90}{\footnotesize{$y~(\mathrm{m})$}}}
      \put(8,4.5){\footnotesize{$0.0$}}
      \put(15,4.5){\footnotesize{$0.8$}}
      \put(22,4.5){\footnotesize{$1.6$}}
       \put(29.5,4.5){\footnotesize{$2.4$}}
      \put(37,4.5){\footnotesize{$3.2$}}
      \put(44,4.5){\footnotesize{$4.0$}}
      \put(56,4.5){\footnotesize{$0.0$}}
      \put(63,4.5){\footnotesize{$0.8$}}
      \put(70,4.5){\footnotesize{$1.6$}}
       \put(77.5,4.5){\footnotesize{$2.4$}}
      \put(85,4.5){\footnotesize{$3.2$}}
      \put(92,4.5){\footnotesize{$4.0$}}
      \put(5,7){\footnotesize{$0.0$}}
      \put(5,14){\footnotesize{$0.8$}}
      \put(5,22){\footnotesize{$1.6$}}
       \put(5,29){\footnotesize{$2.4$}}
      \put(5,36){\footnotesize{$3.2$}}
      \put(5,43){\footnotesize{$4.0$}}
      \end{overpic}
\caption{ 
Evolution of the particle bed (dots) and the pressure fields (contours) in the 2D cylindrical case at (a) $t=2.5~\mathrm{ms}$ and (b) $t=5.0~\mathrm{ms}$ .
}\label{fig:2dcp}	
\end{figure}
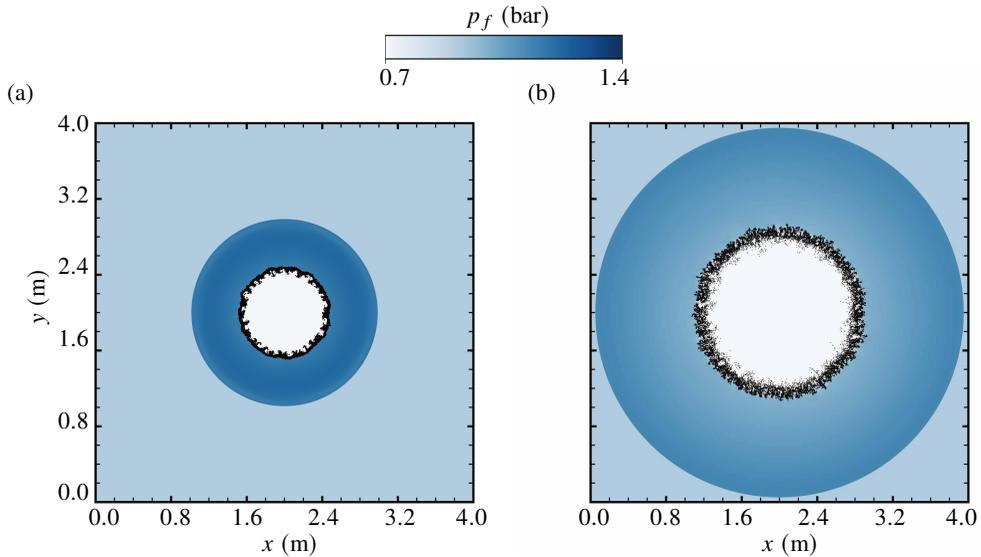

The contour plot of gas pressure and position of the computational parcels are shown in figure~\ref{fig:2dcp}. 
Following the impact an acceleration by high-pressure gas, the particle bed expands radially outward. 
As shown in figure~\ref{fig:2dcp}(a) at $t=2.5~\mathrm{ms}$, internal particle jets emerge first, followed by the appearance of small-scale external particle jets in figure~\ref{fig:2dcp}(b) at $t=5.0~\mathrm{ms}$. 
The number of internal particle jets is larger than that of the external particle jets. 
Contrasted with case DP50, the 2D cylindrical case not only produces distinct internal particle jets but also exhibits a significant delay in the formation of external particle jets.

These computational results are qualitatively consistent with the numerical findings reported by \citet{zhang2015} for liquid dispersion subjected to strong explosive loading in a cylindrical geometry. 
Limited by computational resources, we did not extend the 2D simulation to the late stage. 
Figure 14 of \citet{zhang2015} indicates that at the late stage ($t=100~\mathrm{ms}$), when the gas-particle mixture is under one-way coupled regime, pronounced external particle jets develop.
Hence, the conclusions of the present study, rooted in the drag-dominated mechanism of external particle jet formation under strong explosive loading, are also applicable to cylindrical geometry.

\bibliographystyle{jfm}
\bibliography{jfm}
\end{document}